\documentclass{jfm}
\usepackage{graphicx}
\usepackage{epstopdf, epsfig}
\usepackage{color}

\newcommand{\csr}[1]{{\color{black}#1\normalcolor}}

\shorttitle{Regularized Kirchhoff Rod model}
\shortauthor{N. Ho, K. Leiderman, and S. Olson}

\title{A 3-dimensional model of flagellar swimming in a Brinkman fluid}

\author{NguyenHo Ho\aff{1},
  Karin Leiderman\aff{2},
 \and Sarah Olson\aff{3}\corresp{\email{sdolson@wpi.edu}},}

\affiliation{\aff{1} Department of Mathematical Sciences, University of Cincinnati, 2815 Commons Way, Cincinnati, Ohio 45221, USA.
\aff{2} Applied Mathematics and Statistics, Colorado School of Mines, 1500 Illinois Street, Golden, Colorado 80401 USA.
\aff{3} Department of Mathematical Sciences, Worcester Polytechnic Institute, 100 Institute Road, Worcester, Massachusetts 01609, USA.}

\begin{document}

\maketitle

\begin{abstract}
We investigate 3-dimensional flagellar swimming in a fluid with a sparse network of stationary obstacles or fibers. The Brinkman equation is used to model the average fluid flow where a flow dependent term, including a resistance parameter that is inversely proportional to the permeability, models the resistive effects of the fibers on the fluid. To solve for the local linear and angular velocities that are coupled to the flagellar motion, we extend the method of regularized Brinkmanlets to incorporate a Kirchhoff rod, discretized as point forces and torques along a centerline. 
Representing a flagellum as a Kirchhoff rod, we investigate emergent emergent waveforms for different preferred strain and twist functions.
Since the Kirchhoff rod formulation allows for out-of-plane motion, in addition to studying a preferred planar sine wave configuration, we also study the case with a preferred helical configuration.  Our numerical method is validated by comparing results to asymptotic swimming speeds derived for an infinite-length cylinder propagating planar or helical waves. Similar to the asymptotic analysis for both planar and helical bending, we observe that with small amplitude bending, swimming speed is always enhanced relative to the case with no fibers in the fluid (Stokes) as the resistance parameter is increased. For regimes not accounted for with asymptotic analysis, i.e., large amplitude planar and helical bending, our model results show a non-monotonic change in swimming speed  with respect to the resistance parameter; a maximum swimming speed is observed when the resistance parameter is near one. The non-monotonic behavior is due to the emergent waveforms; as the resistance parameter increases, the swimmer becomes incapable of achieving the amplitude of its preferred configuration. We also show how simulation results of slower swimming speeds for larger resistance parameters are actually consistent with the asymptotic swimming speeds if work in the system is fixed.
\end{abstract}

\begin{keywords}
Authors should not enter keywords on the manuscript, as these must be chosen by the author during the online submission process and will then be added during the typesetting process (see http://journals.cambridge.org/data/\linebreak[3]relatedlink/jfm-\linebreak[3]keywords.pdf for the full list)
\end{keywords}

\section{Introduction}
Microorganisms such as spermatozoa make forward progression by propagating bending along their flagellum. The emergent  flagellar curvature and beat frequency depends on the fluid properties as well as the chemical concentrations within the flagellum \citep{gaffney2011mammalian,Miki07,Smith09b,Suarez06,Woolley01}. The fluid environment experienced by mammalian sperm includes complex geometries and background flows due to interactions with other sperm, cilia, and walls \citep{Fauci06,HoSuarez01,Suarez06,Suarez10b}. As a sperm progresses toward the egg, the fluid could contain differing amounts of uterine cells, sulfomucins, protein networks and other macromolecules, especially at different times in the menstrual cycle   \citep{Katz80,Katz89,Mattner68,Suarez10b}. 
Many experiments have examined sperm motility in gels such as methylcellulose (MC) or polyacrylamide (PA), which may be more representative of the \textit{in vivo} environment. In experiments, the  emergent beat frequency and wavelength varied with viscosity in MC gels \citep{Smith09b} and the swimming speed of mouse sperm decreased in both MC and PA gels (relative to the culture medium) \citep{Suarez92}. This motivates the development of a 3-dimensional (3D) framework to study emergent properties (e.g., waveform or swimming speed) of a sperm when coupled with this protein network. 

Previous computational studies of finite-length swimmers in a Newtonian fluid with preferred bending kinematics have identified that there is a non-monotonic relationship between emergent swimming speeds and bending amplitude \citep{Elgeti10,Fauci95,Olson15}.
On the other hand, for infinite-length flagella with prescribed bending, the asymptotic swimming speeds are an increasing function with respect to the bending amplitude \citep{Taylor51,Taylor52}. Since most gels and biological fluids contain proteins and other macromolecules, recent studies have focused on swimmers in complex fluids. In fluids that exhibit contributions from viscous and elastic effects, the swimming speeds of infinite-length flagella with prescribed kinematics decrease in comparison to the Newtonian case \citep{Fu09,Lauga07}. In contrast, swimming speeds increase for certain parameter choices for a finite-length swimmer in a nonlinear viscoelastic fluid  and a Carreau fluid (Newtonian fluid with shear-dependent viscosity) \citep{Montenegro12,Teran10,Thomases14}. An enhancement in swimming speed relative to a Newtonian fluid has also been observed in a  model of a two-phase fluid for a gel when the elastic network is stationary \citep{Fu10}.

The average fluid flow through an array of sparse, spherical particles can be modeled via the Brinkman equation \citep{Auriault09,Brinkman47,Durlofsky87,Howells74,Spielman}. A flow dependent resistance term accounts for the presence of the particles in the fluid. This type of flow has been studied near boundaries and interfaces \citep{Ahmadi17,Feng}, as well as being a fluid flow to understand flagellar motility of microorganisms. In the case of an infinite-length flagellum in a Brinkman fluid with prescribed bending, in both 2D and 3D, the swimming speed increases as the resistance parameter increases \citep{ho2016swimming,Leshansky09}. This increase in swimming speed is an enhancement relative to the Newtonian case; the presence of particles or fibers actually aids in forward progression.  In contrast,  for a finite-length swimmer with preferred planar bending, there was a non-monotonic relationship between swimming speed and the resistance parameter  \citep{Cortez10,Olson15H,leiderman2016swimming}.

To explore emergent properties of flagellar swimming in  a fluid with a sparse and stationary protein network, we use the incompressible Brinkman equations to govern the fluid motion \citep{Brinkman47,Howells74}:
\begin{eqnarray}
-\nabla p+\mu\Delta {\bf u}-\mu\alpha^2{\bf u}+\mathbf{f}^b&=\mathbf{0},\label{BrEq}\\
\nabla\cdot{\bf u}&=0. \label{BrEqIncomp}
\end{eqnarray}
Here, $p$ is the average fluid pressure (force per area), {\bf u} is the average fluid velocity (length per time),  $\mathbf{f}^b$ represents the body force (force per unit volume) applied on the fluid by the immersed structure, $\mu$ is the viscosity (force time per area), and $\alpha=1/\sqrt{\gamma}$ is the resistance parameter (inverse length), which is assumed constant (isotropic) and inversely proportional to the square root of the permeability $\gamma$. 
One can think of the Brinkman equation as the addition of a lower-order resistance term to the Stokes equations for low Reynolds number flow (since the length scale of these swimmers is small, they live in a viscosity dominated environment where inertia can be neglected). As $\alpha\to 0$ (no resistance),  the incompressible Stokes equations are recovered and as $\alpha\to\infty$ (high resistance), the term $\mu\Delta {\bf u}$ becomes negligible  and Eq.~(\ref{BrEq}) behaves like Darcy's law.  An important characteristic of a Brinkman fluid is the Brinkman screening length, $\sqrt{\gamma}$, which marks the approximate length over which a disturbance to the velocity would decay.

 To consider a microorganism swimming in this environment, we assume that the obstacles are at a low enough volume fraction and far enough apart such that the swimmer is able to easily move between stationary fibers. 
     \begin{figure}
   \hspace{-0.75in}  \textbf{(a)}\hspace{2.25in}\textbf{(b)}\\
 \centering
\includegraphics[height=.35\linewidth]{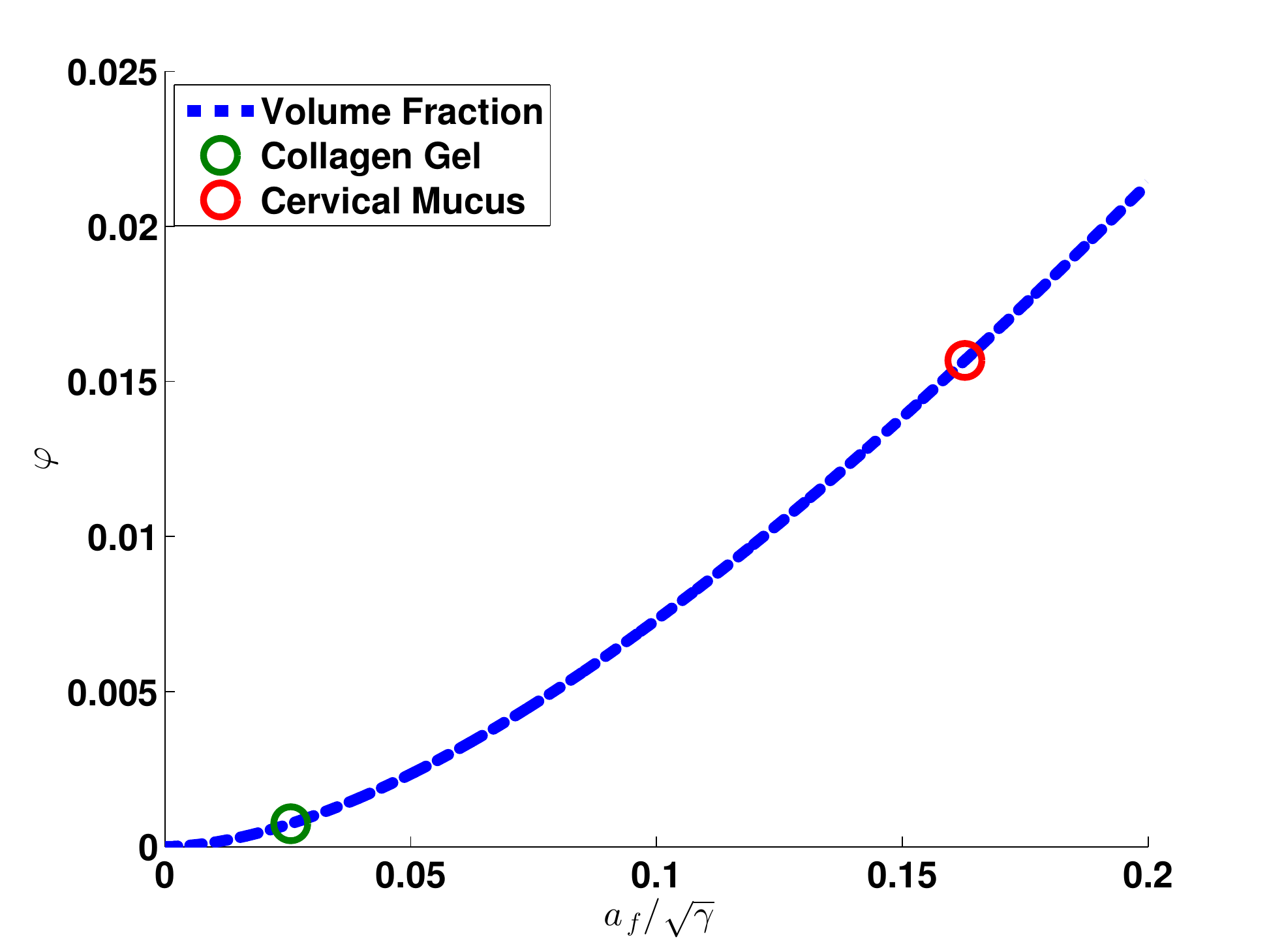}
\includegraphics[height=.35\linewidth]{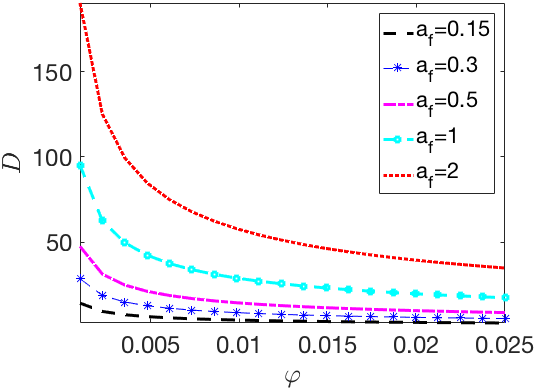}
\caption{\csr{\textit{(a) The volume fraction $\varphi$ solved for in Eq.~(\ref{fibrouseq}) is plotted as a function of $a_f/\sqrt{\gamma}$ where $a_f$ is fiber radius and $\gamma$ is permeability. The green and red markers represent the volume fraction for a collagen gel ($\gamma=8.6$) and cervical mucus ($\gamma=0.0085$), respectively. (b) Using the relevant volume fractions $\varphi$, we plot the interfiber spacing $D$ (in microns) from Eq.~(\ref{ratiointerfiber}) for different fiber radii $a_f$.}}}\label{EstimateGraphPic}
\end{figure}
 In the case of randomly oriented fibers, \citet{Spielman} have derived a relationship between the volume fraction $\varphi$, the permeability $\gamma$, and the radius of the fiber $a_{f}$ as
   \begin{equation}
\frac{a_{f}^2}{\gamma}=4\varphi\left[\frac{1}{3}\frac{a_{f}^2}{\gamma}+\frac{5}{6}\frac{a_{f}}{\sqrt{\gamma}}\frac{K_1(a_{f}/\sqrt{\gamma})}{K_0(a_{f}/\sqrt{\gamma})}\right].\label{fibrouseq}
   \end{equation}
Here, $K_0(\cdot)$ and $K_1(\cdot)$ are the zeroth and first order modified Bessel functions of the second kind.  Figure \ref{EstimateGraphPic} shows the  volume fraction $\varphi$ as a function of the ratio $a_f/\sqrt{\gamma}$, indicating a relevant biological range for $\varphi$. For reference, we also indicate the volume fraction of cervical mucus  and a collagen gel (1 mg/ml) based on experimental values for $\varphi$ and $a_f$ \citep{saltzman1994antibody}. 
The interfiber spacing $D$ (or distance between the fibers) can also be approximated as \citep{Leshansky09}
\begin{equation}
D\approx2a_{f}\left(\frac{1}{2}\sqrt{\frac{3\pi}{\varphi}}-1\right),\label{ratiointerfiber}
\end{equation}
based on a known volume fraction $\varphi$ and fiber radius $a_f$. 
In figure \ref{EstimateGraphPic}(b), we plot the interfiber spacing $D$ for the relevant volume fractions $\varphi$ using a biologically relevant range for $a_f$ \citep{Rutllant01,Rutllant05}. We observe that interfiber spacing is predicted to be in the range of 1-200 microns. In the case of sufficient spacing to allow for a swimmer to move through a fluid with a sparse array of stationary fibers, we assume that the fibers do not impart any additional stress onto the filament. 

A fundamental solution of the incompressible Brinkman equations given in (\ref{BrEq})--(\ref{BrEqIncomp}),  is well-known \citep{Durlofsky87,Pozrikidis1989a}. It represents the velocity due to a concentrated external force acting on the fluid at a single point. However, the velocity becomes singular when the point forces are concentrated along curves in 3D. To eliminate these singular solutions in Stokes flow, the method of regularized Stokeslets \citep{Cortez01,Cortez05} is employed while the method of regularized Brinkmanlets  \citep{Cortez10} is introduced to deal with these situations in a Brinkman fluid.

In order to model emergent waveforms of swimmers that can be either planar or helical, as observed in experiments \citep{Woolley01}, we use a Kirchhoff rod model to represent the elastic flagellum. The propagation of bending along the filament is given as a time-dependent preferred curvature function, where deviations from this preferred configuration lead to forces and moments (body forces). An immersed boundary formulation of the Kirchhoff rod model was first developed by \cite{Lim08} and has been extended to a regularized Stokes formulation \citep{Lee14,Olson13,olson2014motion}. Here, we extend the regularized method to now study flagellar swimming in a fluid governed by the Brinkman equation where we account for the local linear and angular velocity due to point forces and torques along the length of the flagellum. Two approaches for the numerical method are derived in Section \ref{TwoApproaches}, where in the limit as resistance parameter $\alpha\to0$, the solutions approach those of Stokes equations (detailed in Appendix \ref{limits}). We are able to match emergent swimming speeds with asymptotic swimming speeds for both the planar and helical bending cases where swimming speed increases with amplitude and beat frequency for a fixed resistance parameter. The numerical results show that for the planar and helical bending cases, there is an optimal range of $\alpha$, around $\alpha=1$, that allows the swimmer to achieve a large bending amplitude while receiving an extra boost in propulsion from the presence of the fiber network. In addition, as the resistance parameter $\alpha$ increases, the emergent waveform of the swimmer has a decreased amplitude (relative to the preferred amplitude), resulting in a decreased swimming speed.

\section{Kirchhoff Rod Model}\label{BackgroundKR} \label{ChapterKRModel}
With the Kirchhoff Rod (KR) formulation, a flagellum is described by a 3D space curve ${\bf X}(s)$ for $0<s<L$,  where $s$ is a Lagrangian parameter initialized to be the arclength and $L$ is the length of the unstressed rod. Here, we assume the rod length is much greater than the radius and that the rod is isotropic and homogeneous. The associated orthonormal triads $\{{\bf D}^1(s), {\bf D}^2(s), {\bf D}^3(s)\}$ follow the right-handed rule. The triad ${\bf D}^3(s)$ is effectively in the direction of the tangent vector while ${\bf D}^1(s)$ and ${\bf D}^2(s)$ are rotations of the normal and binormal vectors, respectively, coinciding with the principal axes of the rod cross section. 

\begin{figure}
 \centering
\includegraphics[width=.5\linewidth]{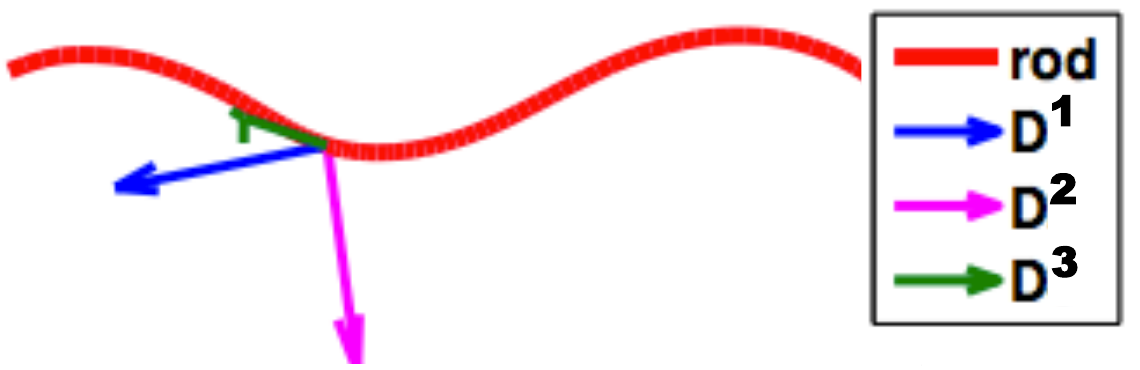}
\caption{The Kirchhoff rod representation of a flagellum is discretized as a helix using a centerline approximation with orthonormal triads \{$\bf{D}^1$, $\bf{D}^2$, $\bf{D}^3$\} plotted at one point on the space curve.}\label{KRmodelPlot}
\end{figure}

Figure \ref{KRmodelPlot} shows the centerline of a flagellum discretized as a helix using the centerline approximation with the associated orthonormal triads plotted at one point on the space curve.  In the standard KR model, ${\bf D}^3(s)$ is enforced to be the tangent vector and the rod is  inextensible. We employ an unconstrained version whereby an elastic energy penalty is used to numerically maintain the inextensibility of the rod and keep ${\bf D}^3(s)$ as a unit tangent vector along the rod \citep{Lim08, Olson13}.

The derivation of the internal force and torque in terms of the associated orthonormal triads has previously been described in detail  \citep{Lim08,Olson13}. Here, we summarize the main equations of the KR model which are utilized later. The balance of force and torque on a cross section of the rod are 
\begin{eqnarray}
0 & = & {\bf f}+\frac{\partial {\bf F}}{\partial s}\label{forceeq1},\\
0 & = & {\bf m}+\frac{\partial {\bf M}}{\partial s}+\left(\frac{\partial {\bf X}}{\partial s}\times {\bf F}\right),\label{torqueeq1}
\end{eqnarray}
where ${\bf f}$ (units of force per length)
and ${\bf m}$ (units of force or torque per length) are part of the external forces applied on the rod. Whereas, ${\bf F}$ and ${\bf M}$ are the internal components of the force transmitted across each section of the rod and are given in terms of ${\bf X}(s)$ and its triads. The components of  ${\bf F}$ and ${\bf M}$ can be expanded in the basis of the triads: 
\begin{eqnarray}
{\bf F}=\sum_{i=1}^{3}F^i{\bf D}^i, \hspace{10mm} {\bf M}=\sum_{i=1}^{3}M^i{\bf D}^i,
\end{eqnarray}
for $i=1, 2, 3$ where both  ${\bf F}$ and ${\bf D}^i$ are 3 by 1 vectors at a given $s$ along the rod centerline. The constitutive relations for the unconstrained KR are  \citep{Olson13,Lim08}
\small{
\begin{equation}
M^1=a_1\left(\frac{\partial {\bf D}^2}{\partial s}\cdot {\bf D}^3-\Omega_1\right), \hspace{1mm} 
M^2=a_2\left(\frac{\partial {\bf D}^3}{\partial s}\cdot {\bf D}^1-\Omega_2\right), \hspace{1mm} 
M^3=a_3\left(\frac{\partial {\bf D}^1}{\partial s}\cdot {\bf D}^2-\Omega_3\right)\label{Nrealtion},
\end{equation}
\begin{equation}
F^1=b_1\frac{\partial {\bf X}}{\partial s}\cdot {\bf D}^1, \hspace{5mm} 
F^2=b_2\frac{\partial {\bf X}}{\partial s}\cdot {\bf D}^2, \hspace{5mm} 
F^3=b_3\left(\frac{\partial {\bf X}}{\partial s}\cdot {\bf D}^3-1\right)\label{Frealtion},
\end{equation}}\noindent where the material properties of the rod are characterized through the parameters $a_i$ and $b_i$ for $i=1,2,3$. The bending moduli are $a_1$, $a_2$ and $a_3$ is the twisting modulus while $b_1, b_2$ are the shear moduli, and $b_3$ is the extensional modulus. The strain-twist vector is represented by $\{\Omega_1, \Omega_2, \Omega_3\}$ where $\Omega_3$ is the intrinsic twist and $\Omega_1, \Omega_2$ are the geodesic and normal curvatures, respectively, associated with the intrinsic curvature $\Omega$ through the equation $\Omega=\sqrt{\Omega_1^2+\Omega_2^2}$. This vector determines the preferred configuration of the rod where internal force and torque are generated by differences from the actual and preferred configuration.   
The preferred strain and twist of the rod can be varied (in time $t$ and with respect to arc length parameter $s$), to propagate planar or helical bending that are representative of sperm flagellar beatforms observed in experiments \citep{Smith09b,Woolley01}.

\section{Method of Regularized Brinkmanlets for the Kirchhoff Rod}
Given a 3D elastic structure immersed in a Brinkman fluid, the equations of motion  include the external forces and torques of the structure on the fluid. The solutions can be calculated exactly in terms of fundamental solutions due to the linearity of the Brinkman equation. However, singular solutions are obtained when evaluating the flow at the location of a point force or torque on the centerline of the structure. Eliminating these singularities requires a regularization method and we utilize the Method of Regularized Brinkmanlets (MRB) \citep{Cortez10}. The idea is to use a smooth approximation to the singular point force or point torque. The smooth approximation is called a ``blob" or regularization function, $\phi_{\varepsilon}(r)$, and is a radially symmetric function 
whose width is determined by the regularization parameter $\varepsilon\ll 1$. In the limit as $\varepsilon\to0$, the singular solutions are recovered.

Since we want to capture both the bending and twisting motions of the rod in a 3D infinite fluid, the expression of the force density ${\bf f}^b$ at a point $\mathbf{x}$ in the fluid is a contribution of both ${\bf f}$ and ${\bf m}$, given as
\begin{equation}
\mathbf{f}^b(\mathbf{x},t)=\int_{\Gamma}\left(-\mathbf{f}(s,t)+\frac{1}{2}\nabla\times(-\mathbf{m}(s,t))\right)\phi_{\varepsilon}(r)ds\label{contfb}
\end{equation}
where $\Gamma$ is the curve corresponding to the centerline of the swimmer and $r=||\mathbf{x}-\mathbf{X}(s,t)||$ \citep{Olson13}.  
In general, we wish to solve the incompressible Brinkman equation in (\ref{BrEq})--(\ref{BrEqIncomp}) with body force as in (\ref{contfb}) for the local linear velocity ${\bf u}$ and the local angular velocity ${\boldsymbol \omega}=\frac{1}{2}\nabla\times{\bf u}$. Once these are known at ${\bf X}(s,t)$, we can update the location of the rod and the associated orthonormal triads using a no-slip condition, 
\begin{eqnarray}
\frac{\partial {\bf X}(s,t)}{\partial t}&=&{\bf u}({\bf X}(s,t)),\label{nosliplocation}\\
 \frac{\partial {\bf D}^i(s,t)}{\partial t}&=&{\boldsymbol \omega}({\bf X}(s,t))\times{\bf D}^i(s,t),\hspace{.3cm}\mbox{$i$=1, 2, 3}.\label{nosliptriads}
 \end{eqnarray}
To simplify the derivation, we focus on a single point force ${\bf f}_c$ and a single point torque  ${\bf m}_c$, both are constants and applied at the point ${\bf X}_c$, given as
 \begin{equation}
{\bf f}^b({\bf x})={\bf f}_c\phi_{\varepsilon}(r)+\frac{1}{2}\nabla\times{\bf m}_c\phi_{\varepsilon}(r)\label{externalforcetorque}
 \end{equation}
 where the blob function $\phi_{\varepsilon}(r)$ has units of inverse volume and $r=\|{\bf x}-{\bf X}_c\|$. We assume that $\phi_{\varepsilon}(r)$ is a radially symmetric function that satisfies the condition $4\pi\int_0^{\infty}r^2\phi_{\varepsilon}(r)dr=1$ in 3D. 
 
\subsection{Pressure Calculation} 
Taking the divergence of both sides of (\ref{BrEq}), where the body force is given by (\ref{externalforcetorque}), results in 
\begin{equation}
  \Delta p=\mu\Delta(\nabla\cdot{\bf u})+\mu\alpha^2 \nabla\cdot {\bf u}+\nabla\cdot({\bf f}_c\phi_{\varepsilon})+\frac{1}{2}\nabla\cdot(\nabla\times{\bf m}_c\phi_{\varepsilon}).\label{DeltapEq}
 \end{equation} 
This is further simplified using the incompressibility condition and vector identities. A radially symmetric $G_{\varepsilon}$ is directly determined by $\Delta G_{\varepsilon}=\phi_{\varepsilon}$ and  
the pressure is thus
    \begin{equation}
   p={\bf f}_c\cdot\nabla G_{\varepsilon}.\label{PressureSolve}
 \end{equation} 
\subsection{Linear and Angular Velocity} 
To find the solution of the linear velocity, we simply substitute (\ref{PressureSolve}) back into (\ref{BrEq}). Given a radially symmetric $G_{\varepsilon}$, one can determine a radially symmetric $B_{\varepsilon}$ where  $(\Delta -\alpha^2)B_{\varepsilon}=G_{\varepsilon}$ for $\alpha=1/\sqrt{\gamma}$. Using  $\Delta B_{\varepsilon}=\alpha^2B_{\varepsilon}+G_{\varepsilon}$, the linear velocity is given as
 \begin{equation}
\mu{\bf u}(\mathbf{x})=({\bf f}_c\cdot\nabla)\nabla B_{\varepsilon}(r)-{\bf f}_c\Delta B_{\varepsilon}(r)-\frac{1}{2}\alpha^2\nabla B_{\varepsilon}(r)\times {\bf m}_c-\frac{1}{2}\nabla G_{\varepsilon}(r)\times {\bf m}_c.\label{linearvel}
 \end{equation} 
On the right hand side of (\ref{linearvel}), the first two terms correspond to the regularized Brinkmanlet due to a point force ${\bf f}_c$. The third and fourth terms are due to a point torque ${\bf m}_c$ and are the regularized Brinkman rotlet.
The angular velocity ${\boldsymbol \omega}$ is then
 \begin{eqnarray}
\mu{\boldsymbol \omega}(\mathbf{x})&=& \frac{1}{2}{\bf f}_c\times\nabla G_{\varepsilon}(r)-\frac{1}{4}({\bf m}_c\cdot\nabla)\nabla G_{\varepsilon}(r)+\frac{1}{4}\Delta G_{\varepsilon}(r){\bf m}_c+\frac{1}{2}\alpha^2{\bf f}_c\times\nabla B_{\varepsilon}(r) \nonumber \\
&&\hspace{1.5in}-\frac{1}{4}\alpha^2({\bf m}_c\cdot\nabla)\nabla B_{\varepsilon}(r)+\frac{1}{4}\alpha^2\Delta B_{\varepsilon}(r) {\bf m}_c.\label{angularvel}
 \end{eqnarray} 
For convenience in evaluating the solutions of (\ref{linearvel}) and (\ref{angularvel}) numerically, we write the local linear and angular velocity  as 
        \begin{eqnarray}
 \mu{\bf u}(\mathbf{x})&=&{\bf f}_cH_1^{\varepsilon}(r)+({\bf f}_c\cdot{\bf \hat{x}}){\bf \hat{x}}H_2^{\varepsilon}(r)+\frac{1}{2}({\bf m}_c\times{\bf \hat{x}})\left[Q_1^{\varepsilon}(r)+\alpha^2Q_2^{\varepsilon}(r)\right],\label{uvel_B}\\
 \mu{\boldsymbol \omega}(\mathbf{x})&=&\frac{1}{2}({\bf f}_c\times{\bf \hat{x}})\left[Q_1^{\varepsilon}(r)+\alpha^2Q_2^{\varepsilon}(r)\right]-\frac{1}{4}\alpha^2\left[{\bf m}_cH_1^{\varepsilon}(r)+({\bf m}_c\cdot{\bf \hat{x}}){\bf \hat{x}}H_2^{\varepsilon}(r)\right]\nonumber\\
&&\hspace{2in}+\frac{1}{4}\left[{\bf m}_cD_1^{\varepsilon}(r)+({\bf m}_c\cdot{\bf \hat{x}}){\bf \hat{x}}D_2^{\varepsilon}(r)\right]\label{wvel_B},
   \end{eqnarray}  
where $\hat{\mathbf{x}}=\mathbf{x}-\mathbf{X}_c$ and $r=||\hat{\mathbf{x}}||$. 
The coefficient functions are given as:
 \begin{eqnarray}
H_1^{\varepsilon}(r)=-\frac{rB''_{\varepsilon}(r)+B'_{\varepsilon}(r)}{r}&,&\hspace{5mm}H_2^{\varepsilon}(r)=\frac{rB''_{\varepsilon}(r)-B'_{\varepsilon}(r)}{r^3},\label{hgeneral}\\
Q_{1}^{\varepsilon}(r)=\frac{G'_{\varepsilon}(r)}{r}&,&\hspace{5mm}Q^{\varepsilon}_{2}(r)=\frac{B'_{\varepsilon}(r)}{r},\label{qgeneral} \\
 D_{1}^{\varepsilon}(r)=\phi_{\varepsilon}(r)-Q^{\varepsilon}_1(r)&,&\hspace{5mm}D^{\varepsilon}_2(r)=-\frac{rG''_{\varepsilon}(r)-G'_{\varepsilon}(r)}{r^3},\label{dgeneral}
 \end{eqnarray} 
and additional details on the derivation are given  in Appendix \ref{CoefficientsFunctions} and Section \ref{TwoApproaches}.

\subsection{Regularized Coefficient Functions} \label{TwoApproaches}
Previously, \cite{Cortez10} detailed two different approaches that can be used to determine appropriate coefficient functions $H_i^{\varepsilon}$ for $i=1,2$ in the case of 3D Brinkman flow due to regularized point forces (point torques were not considered). The approaches are to either first start with an appropriate blob function $\phi_{\varepsilon}(r)$ or to first start by choosing a regularization of the well known singular solution; each approach leads to a slightly different PDE, resulting in different coefficient functions $H_i^{\varepsilon}$. On the test cases of flow past a stationary cylinder (2D) or sphere (3D), the error does depend on both the regularization parameter $\varepsilon$ and resistance parameter $\alpha$ \citep{Cortez10,leiderman2016swimming}. We note that each approach leads to a slightly different regularization of the forces, which in turn, results in a slightly different flow. Carefully chosen blob functions can reduce error and allow for computationally efficient expressions to calculate regularized Stokes and Brinkman flow \citep{Nguyen13,NguyenKarinTriply}. On the other hand, a particular regularization of the singular solutions may also lead to desired properties. The best method to use will depend on the application and the desired results in terms of error and computational ease. Hence, we describe how to find the regularized coefficient functions $H_i^{\varepsilon}$, $D_i^{\varepsilon}$, and $Q_i^{\varepsilon}$ for $i=1,2$ using both approaches. In Appendix \ref{appcompare}, we compare the blob functions and simulation results for the two approaches. Additionally, in Appendix \ref{MRBrinman1}$-$\ref{MRBrinman2} we show that when $\alpha\to0$, in both cases, the solution for the linear and angular velocity in (\ref{uvel_B})--(\ref{wvel_B}) approaches  the corresponding regularized solution for Stokes flow \citep{Olson13}. 

\subsubsection{Option 1: Regularizing the Fundamental Solutions}\label{KRRegularizedApproach}
The first approach is to solve for the pressure, as well as the linear and angular velocity given in (\ref{linearvel})--(\ref{angularvel}) by regularizing the fundamental solutions \citep{Cortez10}. That is, the functions $G_{\varepsilon}(r)$ and $B_{\varepsilon}(r)$ are written as 
   \begin{equation}
G_{\varepsilon}(r)=-\frac{1}{4\pi R},\hspace{10mm} B_{\varepsilon}(r)=\frac{1-e^{-\alpha R}}{4\pi\alpha^2R},\label{fundreg}
 \end{equation}
with  $R=\sqrt{r^2+\varepsilon^2}$ such that when $\varepsilon\rightarrow 0$, we recover the singular solutions. The corresponding regularized solutions of $H_i^{\varepsilon}(r), Q_i^{\varepsilon}(r)$, and $D_i^{\varepsilon}(r)$ for $i=1, 2$ using (\ref{hgeneral})--(\ref{dgeneral}) and (\ref{fundreg}) are detailed in Appendix \ref{option1app}.
 
 \subsubsection{Option 2: Choosing a Blob Function}\label{KRBlobApproach}
The second approach, which we use for the results in Section \ref{TestCaseSection}, is to first choose a suitable blob function and then derive the corresponding fundamental solutions. We can determine $G_{\varepsilon}(r)$ and $B_{\varepsilon}(r)$ for a given blob function $\phi_{\varepsilon}(r)$ as:
  \begin{eqnarray*}
G_{\varepsilon}(r)&=&-\int_{0}^{\infty}t\phi_{\varepsilon}(t)dt+\frac{1}{r}\int_0^r(r-t)t\phi_{\varepsilon}(t)dt,\\
B_{\varepsilon}(r)&=&\frac{1}{\alpha^2}\int_{0}^{\infty}\left[1-\frac{\sinh(\alpha r)}{\alpha r}e^{-\alpha t}\right]t \phi_{\varepsilon}(t)dt+\frac{1}{\alpha^3 r}\int_0^{r}\left[\sinh(\alpha(r-t))-\alpha(r-t)\right]t\phi_{\varepsilon}(t)dt.
 \end{eqnarray*} 
We utilize the 3D blob function
\begin{equation}
\phi_{\varepsilon}(r)=(a_0+a_1r^2)e^{-r^2/\varepsilon^2},\label{NganBlobFunction}
 \end{equation} %
where the coefficients $a_0$ and $a_1$ are 
     \begin{eqnarray*}
a_0=\frac{1}{\alpha^2\varepsilon^5\pi^{3/2}}\left(\alpha^2\varepsilon^2+6-6e^{-\alpha^2\varepsilon^2/4}\right), \hspace{5mm}
a_1=-\frac{4}{\alpha^2\varepsilon^7\pi^{3/2}}\left(1-e^{-\alpha^2\varepsilon^2/4}\right).
 \end{eqnarray*}
This blob function was previously derived for a triply periodic Brinkman fluid \citep{NguyenKarinTriply}.
 The resulting functions $G_{\varepsilon}(r)$ and $B_{\varepsilon}(r)$ are

   \begin{eqnarray}
G_{\varepsilon}(r)&=&-\frac{1}{\pi ^{3/2}\alpha^2 \varepsilon^3}\left(e^{\alpha^2 \varepsilon^2/4}-1\right)e^{-\alpha^2 \varepsilon^2/4-r^2/\varepsilon^2}-\frac{1}{4\pi  r}\textrm{erf}\left(\frac{r}{\varepsilon}\right),\label{gblob}\\
B_{\varepsilon}(r)&=&\frac{1}{8\pi \alpha^2 r}\left[2-2\textrm{erfc}\left(\frac{r}{\varepsilon}\right)-e^{-\alpha r}\textrm{erfc}\left(\frac{\alpha\varepsilon}{2}-\frac{r}{\varepsilon}\right)+e^{\alpha r}\textrm{erfc}\left(\frac{\alpha\varepsilon}{2}+\frac{r}{\varepsilon}\right)\right].\label{bblob}
 \end{eqnarray} 
The equations for $H_i^{\varepsilon},~ Q_i^{\varepsilon},~ D_i^{\varepsilon}$ for $i=1,2$ using (\ref{NganBlobFunction}) can be derived from (\ref{hgeneral})--(\ref{dgeneral}) and are given in Appendix \ref{option2app}.   

     \section{Numerical Algorithm}\label{AlgorithmKR}
\indent The algorithm for calculating the local fluid flow given in (\ref{uvel_B})--(\ref{wvel_B}), as well as the procedure for updating the configuration of the rod through the no-slip boundary conditions in (\ref{nosliplocation})--(\ref{nosliptriads})  is similar to previous derivations \citep{Lim08, Lim10, Olson13}, except that we are now solving the Brinkman equation and have additional terms as described in the previous sections. The centerline of the rod is discretized into $N$ immersed boundary points where $s_k=k\triangle s$, for $k=1, \dots, N$ and $\triangle s$ is a fixed uniform spacing. Let $n$ be the time-step index for time $t=n\triangle t$ where $\triangle t$ is the time step. Then, ${\bf u}^n_k={\bf u}({\bf X}_k^n)={\bf u}({\bf X}(k\triangle s,n\triangle t))$ is the fluid velocity at time step $n$ at rod location $s_k$. 

Assuming a given configuration of the rod (centerline and orthonormal triad) with associated material and intrinsic parameters $a_i$, $b_i$, $\Omega_i$ for $i=1,2,3$, the numerical algorithm is as follows:
\begin{enumerate}
\item  Evaluate the orthonormal triads at half grid points $s_{k+1/2}$ through the use of an orthogonal rotation matrix. Let $\mathcal{D}_k=[\mathbf{D}^1_k\hspace{.1cm} \mathbf{D}^2_k\hspace{.1cm} \mathbf{D}^3_k]$ where $\mathbf{D}^i_k$ is a 3 by 1 vector at  time step $n$, so that $\mathcal{D}_k^T\mathcal{D}_k$ is the identity matrix and $\mathcal{D}_{k+1}=\mathcal{A}\mathcal{D}_k$ where $\mathcal{A}=\mathcal{D}_{k+1}\mathcal{D}_k^T$ is the rotation matrix mapping ${\bf D}_k^i$ to ${\bf D}_{k+1}^i$ for $i=1,2,3$. The interpolation to $k+1/2$ is performed by applying the principal square root of $\mathcal{A}$ as follows: 
\begin{equation} 
\mathcal{D}_{k+1/2}=\sqrt{\mathcal{A}}\mathcal{D}_k.
\end{equation}
Here, $\sqrt{\mathcal{A}}$ is a rotation about the same axis as $\mathcal{A}$ but by half the angle.
\item Using the updated triad at the half grid point at time step $n$, the internal force and internal moment transmitted across the cross section of the rod is evaluated at $s_{k+1/2}$ using \begin{eqnarray}
M^i_{k+\frac{1}{2}}&=&a_i\left(\frac{{\bf D}_{k+1}^j-{\bf D}_{k}^j}{\triangle s}\cdot {\bf D}^j_{k+\frac{1}{2}}-\Omega_i\right)\label{momentmethod1},\\
F^i_{k+\frac{1}{2}}&=&b_i\left(\frac{{\bf X}_{k+1}-{\bf X}_{k}}{\triangle s}\cdot {\bf D}^i_{k+\frac{1}{2}}-\delta_{3i}\right)\label{forcemethod1},
\end{eqnarray}
where $\delta_{3i}$ is the Kronecker delta. The force ${\bf F}_{k+1/2}$ and moment ${\bf M}_{k+1/2}$ vectors are assembled as
\begin{equation}
{\bf F}_{k+\frac{1}{2}}=\sum_{i=1}^{3}F_{k+\frac{1}{2}}^i{\bf D}_{k+\frac{1}{2}}^i,\hspace{15mm}{\bf M}_{k+\frac{1}{2}}=\sum_{i=1}^{3}M_{k+\frac{1}{2}}^i{\bf D}_{k+\frac{1}{2}}^i.
\end{equation}
The force $\mathbf{f}$ and torque $\mathbf{m}$ exerted on the fluid by the rod from (\ref{forceeq1})--(\ref{torqueeq1}) are discretized using a standard central difference to determine $\mathbf{f}_k$ and $\mathbf{m}_k$ for $k=1,\ldots,N$.
\item 
The body force in (\ref{contfb}) is approximated as 
\begin{equation}
\mathbf{f}^b=\sum_{k=1}^N(-\mathbf{f}_k\triangle s)\phi_{\varepsilon}+\frac{1}{2}\sum_{k=1}^N(-\nabla\times \mathbf{m}_k\triangle s)\phi_{\varepsilon}.\label{thrustforce}
\end{equation}
With $N$ point forces and torques applied on the fluid, 
by superposition 
the linear and angular velocities of the fluid at any point ${\bf x}$ are calculated as 
\begin{footnotesize}
     \begin{eqnarray}
 \mu{\bf u}(\mathbf{x})&=&\sum_{k=1}^N-\left[({\bf f}_k\triangle sH_1^{\varepsilon}(r)+({\bf f}_k\triangle s\cdot{\bf \hat{x}}){\bf \hat{x}}H_2^{\varepsilon}(r)+\frac{1}{2}({\bf m}_k\triangle s\times{\bf \hat{x}})\left[Q_1^{\varepsilon}(r)+\alpha^2Q_2^{\varepsilon}(r)\right]\right],\label{uvel_Br}\\
 \mu{\boldsymbol \omega}(\mathbf{x})&=&\sum_{k=1}^N\left[-\frac{1}{2}({\bf f}_k\triangle s\times{\bf \hat{x}})\left[Q_1^{\varepsilon}(r)+\alpha^2Q_2^{\varepsilon}(r)\right]+\frac{1}{4}\alpha^2\left[{\bf m}_k\triangle sH_1^{\varepsilon}(r)+({\bf m}_k\triangle s\cdot{\bf \hat{x}}){\bf \hat{x}}H_2^{\varepsilon}(r)\right]\right]\nonumber\\
&&\hspace{1.5in}-\sum_{k=1}^N\left[\frac{1}{4}\left[{\bf m}_k\triangle sD_1^{\varepsilon}(r)+({\bf m}_k\triangle s\cdot{\bf \hat{x}}){\bf \hat{x}}D_2^{\varepsilon}(r)\right]\right]\label{wvel_Br},
   \end{eqnarray}  

 \end{footnotesize}\noindent
for $\hat{\mathbf{x}}={\bf x}-{\bf X}_k$ and $r=\|\hat{\bf x}\|$.
\item Next, to update the position of the rod, we use the no-slip boundary condition in (\ref{nosliplocation}) written in terms of the  Euler method as
   \begin{equation}
{\bf X}_k^{n+1}={\bf X}^n_k+{\bf u}({\bf X}_k^n)\triangle t.\label{LocaltionKR}
   \end{equation}
The orthonormal triads are updated through (\ref{nosliptriads})    using
      \begin{equation}
(\mathcal{D}_k)^{n+1}={\bf \mathcal{R}}\left(\frac{{\boldsymbol \omega}({\bf X}^n_k)}{\|{\boldsymbol \omega}({\bf X}^n_k)\|},\|{\boldsymbol \omega}({\bf X}^n_k)\|\triangle t\right)(\mathcal{D}_k)^n,\label{TriadKR}
   \end{equation}
where ${\bf \mathcal{R}}({\bf e}, \theta)$ is an orthogonal matrix rotating around an angle $\theta$ about the axis of the unit vector ${\bf e}$ and is defined as
\begin{eqnarray*}
{\bf \mathcal{R}}({\bf e}, \theta)=(\cos\theta) {\bf I}+(1-\cos\theta){\bf e}{\bf e}^T+\sin\theta({\bf e}\times),
\end{eqnarray*}
where ${\bf I}$ is the 3 by 3 identity matrix and ${\bf e}\times$ is a 3 by 3 antisymmetric matrix. The matrix ${\bf \mathcal{R}}({\bf e}, \theta)$ is often called the Rodrigues rotation matrix \citep{Crisfield97}. When determining the new location ${\bf X}_k^{n+1}$, higher order methods can be used to update ${\bf X}_k^{n+1}$, e.g. second order or fourth order Runge-Kutta methods.
\end{enumerate}
\hspace{0.5cm}We note that the regularization parameter $\varepsilon$ should be chosen carefully since it is both a numerical parameter and a physical parameter.  In the limit as $\varepsilon\to0$, we approach the singular solution and thus, regularization error will depend on the choice of $\varepsilon$. In addition, it can be considered a physical parameter since it controls the width of the region where the force is spread. In this KR model, the equations for force and torque balance are derived under the assumption that the rod radius is much smaller than the length of the rod. Thus, we want to ensure that the force is spread to a region of similar radius to that of the rod or filament. Previous studies have also looked at how the error for the method of regularized Brinkmanlets varies as both $\varepsilon$ and $\alpha$ vary \citep{Cortez10,leiderman2016swimming}. For our application, we choose a fixed regularization parameter $\varepsilon$ for all simulations that matches well with asymptotic results and that lends itself to a range corresponding to the physical radius of the flagellum.

\section{Results}\label{TestCaseSection}
We study the effect of varying the resistance parameter $\alpha$ on the behavior and overall performance of swimmers propagating planar or helical bending. 
The time step used is $\triangle t=10^{-6}$ s and the fluid viscosity is set to $\mu=10^{-6}~g$ $\mu m^{-1}$s$^{-1}$ (viscosity of water at room temperature). The parameter values and units are summarized in Table \ref{UnitTable}. In all of our test cases, a force and moment free boundary condition is prescribed, corresponding to 
$${\bf F}_{1/2}=\bf{F}_{N+1/2}=0, \hspace{2em} {\bf M}_{1/2}={\bf M}_{N+1/2}=0.$$
\begin{table}
\centering
\begin{tabular}{c|c|ccccc}
\hline\hline
Parameters & Symbols& Units\\
\hline
Points along the rod & N & \\
Length& $L$ & $\mu$m\\
Wavelength, Amplitude, Radius & $\lambda, b, r$ & $\mu$m\\
Mesh width of the rod& $\triangle s=L/(N-1)$ & $\mu$m\\
Regularization parameter& $\varepsilon$& $\mu$m \\
Resistance & $\alpha$ & $\mu$m$^{-1}$\\
Time & $t$ & seconds (s)\\
Beat Frequency $f$ & $\sigma/2\pi$& $Hz$ (1/s)\\
 \hline
\end{tabular}
\caption{Table of parameters (with units) used in Section \ref{TestCaseSection}.}\label{UnitTable}
\end{table}

\subsection{Planar Bending}\label{planarsection}
The Regularized KR method is validated by comparing the numerical and asymptotic swimming speed of a cylindrical tail propagating small amplitude planar bending. In simulations, the KR is initialized as a straight rod with orthonormal triad $\mathbf{D}^1(s)=(1,0,0)$, $\mathbf{D}^2(s)=(0,\cos(\zeta_p),-\sin(\zeta_p))$, and $\mathbf{D}^3(s)=(0,\sin(\zeta_p),\cos(\zeta_p))$ for a small perturbation $\zeta_p=0.001$ to ensure the rod is not initialized in an equilibrium configuration \citep{Lim10}. The rod is given the following preferred strain and twist, 
\begin{eqnarray*}
\Omega_1=-bk^2\sin(ks+\sigma t), \hspace{2mm} \Omega_2=0, \hspace{2mm} \Omega_3=0,\label{pcurva}
 \end{eqnarray*} 
which corresponds to a swimmer trying to achieve the preferred configuration of a sinusoidal waveform parameterized by  $x(s,t)=0$, $y(s,t)=b\sin(ks+\sigma t)$, and $z(s,t)=s$ \citep{Olson13}.
Here, the rod is bending in the $y$-plane and the wave propagates along the filament in the $z$ direction. The bending amplitude is $b$, $f=\sigma/2\pi$ is the beat frequency, $U=\sigma/k$ is the velocity of the propagating wave where $\sigma$ is the frequency of the wave, and the wavenumber is $k=2\pi/\lambda$ where $\lambda$ is the wavelength. Since this is a preferred curvature model, the achieved bending amplitude and the swimming speed  are emergent properties of the coupled system.

We previously derived the asymptotic swimming speed  for a cylinder of infinite-length propagating small amplitude planar bending \citep{ho2016swimming}. This is calculated as 
 \begin{eqnarray}
U_{\infty}=\frac{1}{2}b^2k\sigma\left[\frac{(1-\chi^2)K_0(\zeta_1)+\chi^2\log\chi}{(1-\chi^2)K_0(\zeta_1)-(2-\chi^2)\log\chi}\right].\label{asymp}
 \end{eqnarray}
Here, $K_0(\cdot)$ is the zeroth order modified Bessel function of the second kind, $\chi=\sqrt{1+\alpha^2/k^2}$, and $\zeta_1=kr_a\ll1$ where $r_a$ is the radius of the rod. To compare this result to those from the simulations, we set $r_a=4.76\triangle s$.
  \begin{figure}
\hspace{1.1cm}{\bf(a)}\hspace{5.5cm}{\bf(b)}\\ \vspace{-.3cm}
 \begin{center}
\includegraphics[width=.43\linewidth]{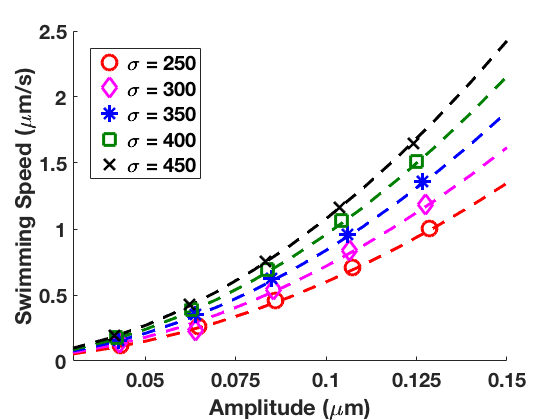}
\includegraphics[height=.32 \linewidth]{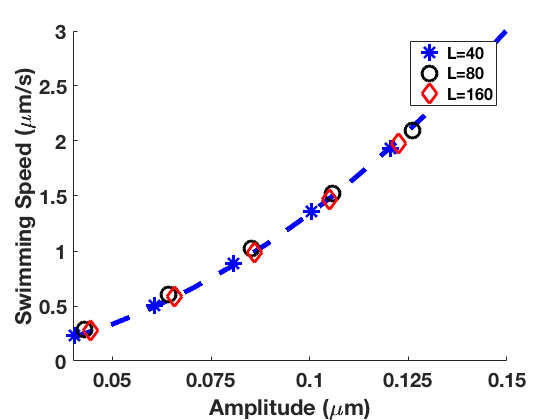}\\
\end{center}
\caption{Comparison of swimming speeds for planar bending using resistance parameter $\alpha=0.1$ where symbols are numerical results  and the dashed lines are the asymptotic swimming speed. {(a)}  Results for various beat frequencies $f=\sigma/2\pi$  {using a rod length of $L=40$ (600  points on the swimmer)}. {(b) The swimming speeds of filaments with lengths $L=40, L=80$, and $L=160$ for {$\sigma=350$}}. For both (a)--(b), stiffness coefficients used are reported in the SIMS1 column of Table \ref{tablecoefficientsWaveForms} and $\varepsilon=6.363\triangle s$.}\label{L40L80}
\end{figure}

In figure \ref{L40L80}(a), we compare the swimming speed results for resistance parameter $\alpha=0.1$. The computational swimming speed for the finite-length swimmer is shown with symbols and is determined as the average swimming speed in the $z$-direction along the length of the rod. The asymptotic swimming speed of the infinite-length swimmer is given by $U_{\infty}$ in (\ref{asymp}) and corresponds to the dashed lines. For different beat frequencies $f=\sigma/2\pi$ with $\sigma$ in the range of 250 to 500, the swimming speeds from the computational method in figure \ref{L40L80}(a) scale quadratically with respect to the amplitude $b$ (for small amplitude with $b\leq0.15$), following the trend of the asymptotic swimming speeds. In the simulations, we are assigning a preferred curvature and hence a preferred amplitude $b$ of either 0.05, 0.075, 0.1, 0.125, or 0.15.  The numerical swimming speeds are plotted in figure \ref{L40L80}(a)--(b) with respect to the achieved amplitude (at $t=0.6$ s); achieved amplitudes are smaller than the preferred amplitudes for this parameter set but the achieved amplitude increases as beat frequency $f=\sigma/2\pi$ decreases (shown in figure \ref{L40L80}(a)) since there is more time for the swimmer to reach the preferred amplitude. We observe excellent agreement between the asymptotic and numerical swimming speeds for a range of $\alpha$, and swimming speeds increase as $\alpha$ increases in the case of small preferred amplitude $b$ (results not shown). 

Since the asymptotic swimming speed in (\ref{asymp}) is for an infinite-length swimmer, we explore the emergent swimming speed for different filament lengths in figure \ref{L40L80}(b). Here, with $\sigma=350$, the longest finite-length swimmer at $L=160$ has the best agreement with the asymptotics. Interestingly, the $L=80$ finite-length swimmer achieves slightly faster swimming speeds than the $L=40$ and $L=160$ swimmer for this particular parameter set. We note that since the derivation of the asymptotic swimming speed for the infinite-length swimmer assumed small radius and amplitude, as expected, the asymptotics vary from the numerical simulations for amplitude $b>0.15$ and the asymptotics fail for a radius $r_a\geq 1$. The KR model assumes that the length is much greater than the radius, thus we can simulate rods with this radius when rod length $L>100$. 

\begin{table}
\centering
\begin{tabular}{c|c|c|cccc}
\hline\hline
  Parameters & SIMS1& SIMS2 & SIMS3      \\
\hline
Bending modulus, $a=a_1=a_2$ (g $\mu$m$^3$ s$^{-2}$)&3.5 $\times$ $10^{-3}$&0.1&1\\
Twisting modulus, $a_3$ (g $\mu$m$^3$ s$^{-2}$)&3.5 $\times$ $10^{-3}$&0.1&1\\
Shearing modulus, $b=b_1=b_2$ (g $\mu$m$ $ s$^{-2}$)&8.0 $\times$ $10^{-1}$&0.06&0.6\\
Stretching modulus, $b_3$ (g $\mu$m$ $ s$^{-2}$)&8.0 $\times$ $10^{-1}$&0.06&0.6\\
 \hline
\end{tabular}
\caption{Stiffness coefficients for the different simulations.}\label{tablecoefficientsWaveForms}
\end{table}
 
Next, we investigate emergent waveforms and swimming speeds that result from varying the resistance parameter $\alpha$. 
The swimmers are initialized as straight rods with a preferred planar curvature as in (\ref{pcurva}).
The sperm flagellum is represented as a centerline, which we set to $L=50$ (discretized with 301 points). The range of parameters for the preferred waveform are based on previous experiments in different fluid environments where beat frequency ranges from 10--20 Hz,  wavelength is in the range of 10--60$~\mu$m, and the mean amplitude can be as large as $6~\mu$m \citep{Smith09b,vernon1999three}. The stiffness values in the SIMS2 column of Table \ref{tablecoefficientsWaveForms} correspond to experimental values estimated for interdoublet bending resistance and shear resistance in sea urchin sperm  \citep{pelle2009mechanical}. 
\begin{figure}
 \hspace{0.65in}{\bf (a)}\\ \vspace{-.3cm}
 \begin{center}
\includegraphics[height=.25\linewidth,width=0.7\linewidth]{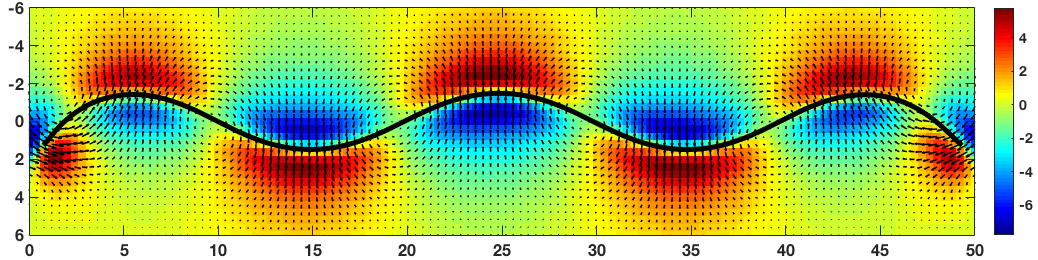}\\
\end{center}
 \hspace{0.65in}{\bf (b)}\\ \vspace{-.3cm}
\begin{center}
\includegraphics[height=.25\linewidth,width=0.71\linewidth]{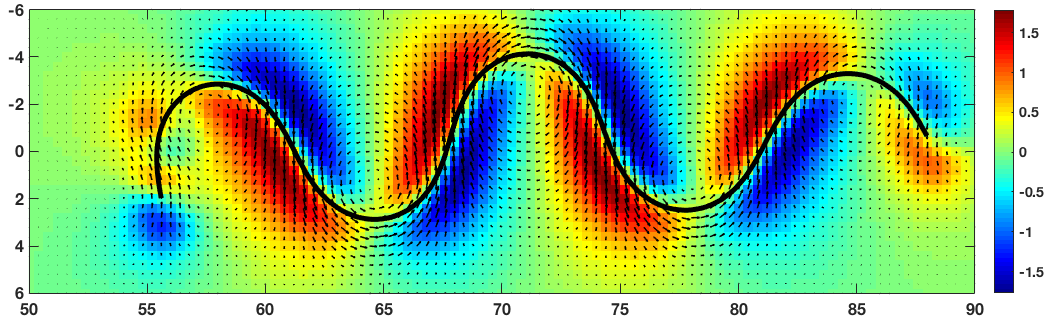}
\end{center}
\caption{The velocity field (black arrows) and the pressure (colorbar to the right, scaled by $p^*=10^4$) of a swimmer with planar bending in a Brinkman fluid with resistance parameter $\alpha=1$ in the plane $x=0$ at (a) $t=0.0012$ s and at (b) $t=1.2$ s. Parameters used are: amplitude $b=4$, wavelength $\lambda=20$, regularization parameter $\varepsilon=5.454\triangle s$, $f=20$, and stiffness coefficients correspond to SIMS2 in Table  \ref{tablecoefficientsWaveForms}.}\label{PressVel_Planar}
\end{figure}

Figure \ref{PressVel_Planar} shows an example of the velocity field and pressure in the plane $x=0$ around a swimmer propagating planar bending. Due to forces and torques along the length, local vortices of flow are observed, changing direction with the local concavity of the swimmer. Since the swimmer is initialized to be straight, we observe in figure \ref{PressVel_Planar}(a)  at $t=0.0012$ s that there is a sinusoidal wave but it has not yet achieved the preferred amplitude and, in (b), the swimmer achieves the preferred amplitude by $t=1.2$ s.   
   \begin{figure}
   \hspace{1cm}{\bf(a)}\hspace{4.9cm}{\bf(b)}\\ \vspace{-.3cm}
 \begin{center}
\includegraphics[height=.28\linewidth,width=0.46\linewidth]{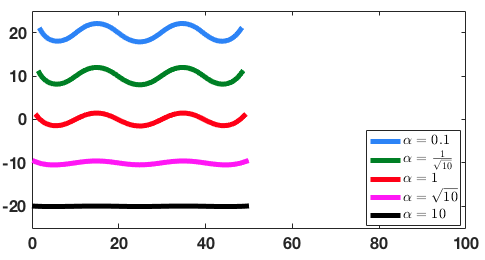}
\includegraphics[height=.28\linewidth,width=0.45\linewidth]{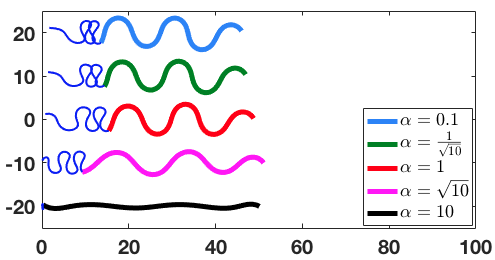}\\
\end{center}
 \hspace{1.1cm}{\bf(c)}\hspace{4.9cm}{\bf(d)}\\ \vspace{-.3cm}
 \begin{center}
\includegraphics[height=.28\linewidth,width=0.45\linewidth]{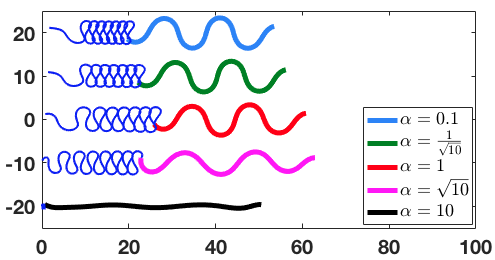}
\includegraphics[height=.29\linewidth,width=0.46\linewidth]{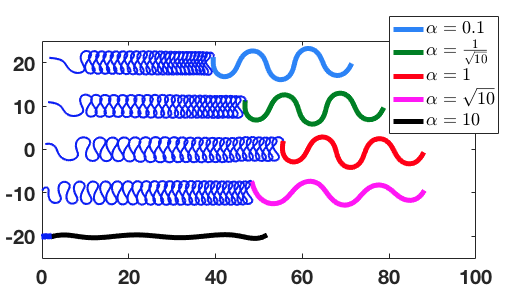}\\
\end{center}
\caption{Snapshots of swimmers in a Brinkman fluid with different resistance parameter $\alpha$ at (a) $t=0.0012$ s, (b) $t=0.12$ s, (c) $t=0.6$ s and (d) $t=1.2$ s. The endpoints of the swimmers are also tracked and plotted with a blue line. Parameters used are: amplitude $b=4$, wavelength $\lambda=20$, regularization parameter $\varepsilon=5.454\triangle s$, $f=20$, and stiffness coefficients correspond to SIMS2 in Table \ref{tablecoefficientsWaveForms}.}\label{SnapShot1}
\end{figure}

As the resistance parameter $\alpha$ is varied, we observe a non-monotonic change in swimming speed for this same fixed set of parameters. 
Four different snapshots of the swimmers in time are shown in figure \ref{SnapShot1} in the plane $x=0$. At each time point, the five different swimmers correspond to a different simulation at a particular $\alpha$; they are placed on the same figure for comparison. The same preferred curvature function (with the same amplitude $b$) is used for each simulation. To monitor the trajectory of the swimmers, the endpoint is also tracked and plotted. At $t=0.0012$ s, shown in figure \ref{SnapShot1}(a), the swimmers immersed in a fluid with smaller resistance parameter (e.g. $\alpha=0.1, 1/\sqrt{10}, 1$) deform and interact with the fluid to start propagating sinusoidal bending while the swimmer in the fluid with $\alpha=10$ remains in a fairly straight line. Figure \ref{SnapShot1}(b) shows that the cases with smaller $\alpha$ have swimmers making forward progression. The swimmers in a fluid with resistance parameter $\alpha=0.1$, $1/{\sqrt{10}}$, and 1 are able to achieve an amplitude close to $b=4$. In contrast, in the case of $\alpha= \sqrt{10}$, the swimmer is not able to reach the desired amplitude and for the $\alpha=10$ simulation, the swimmer is barely moving. This phenomenon occurs because the additional fluid resistance is preventing the swimmers from generating the preferred amplitude. One can imagine that either the fiber radius or the volume fraction of fibers increases as the resistance parameter increases. Thus, the presence of the fibers greatly hinders the ability of the swimmer to generate the preferred waveform. 

At time $t=0.6$ s, figure \ref{SnapShot1}(c) shows that the fastest swimming occurs in the fluid with resistance parameter $\alpha=\sqrt{10}$, where the swimmer's emergent waveform has a longer wavelength and smaller amplitude than the preferred ones. 
By time $t=1.2$ s, shown in figure \ref{SnapShot1}(d), the fastest swimmers are in a fluid with $\alpha=\sqrt{10}$ and $\alpha=1$, the next fastest is $\alpha=1/\sqrt{10}$, and then $\alpha=0.1$.  Tracking the last point of each flagellum in figure \ref{SnapShot1}, we observe that each swimmer exhibits a figure-eight motion in the plane of swimming. Greater propulsion in the $z$-direction with each beat of the tail of the swimmer is observed for $\alpha=\sqrt{10}$ and 1. 
  
  \begin{figure}
 \hspace{0.85cm}{\bf (a)} \hspace{16.5em}{\bf (b)}\\ \vspace{-0.3cm}
  \begin{center}
\includegraphics[height=.3\linewidth,width=0.49\linewidth]{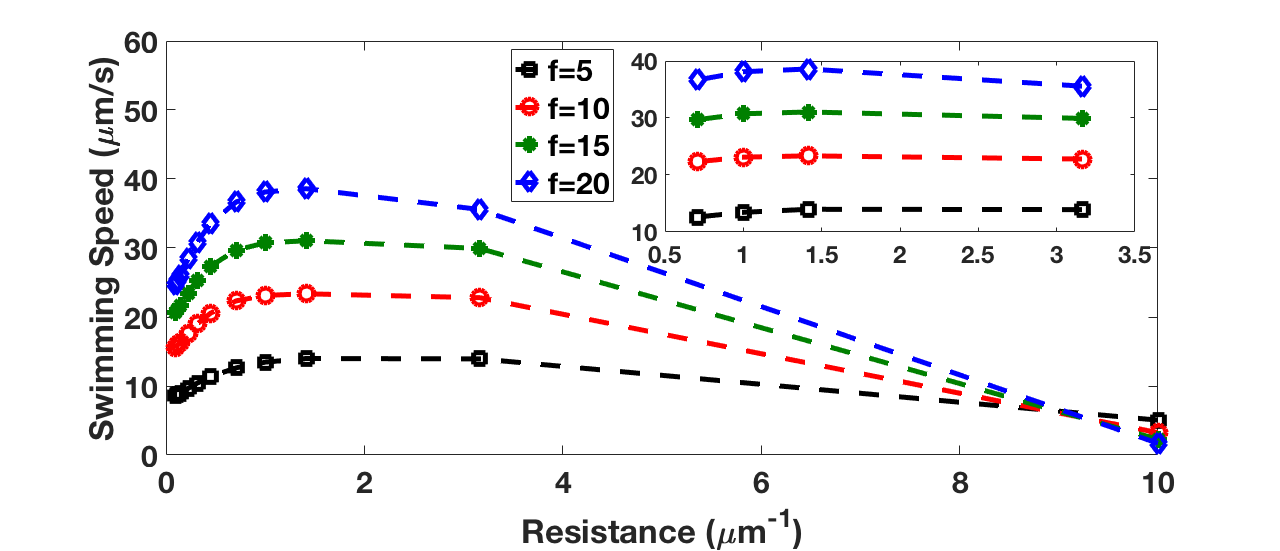}
\includegraphics[height=.3\linewidth,width=0.49\linewidth]{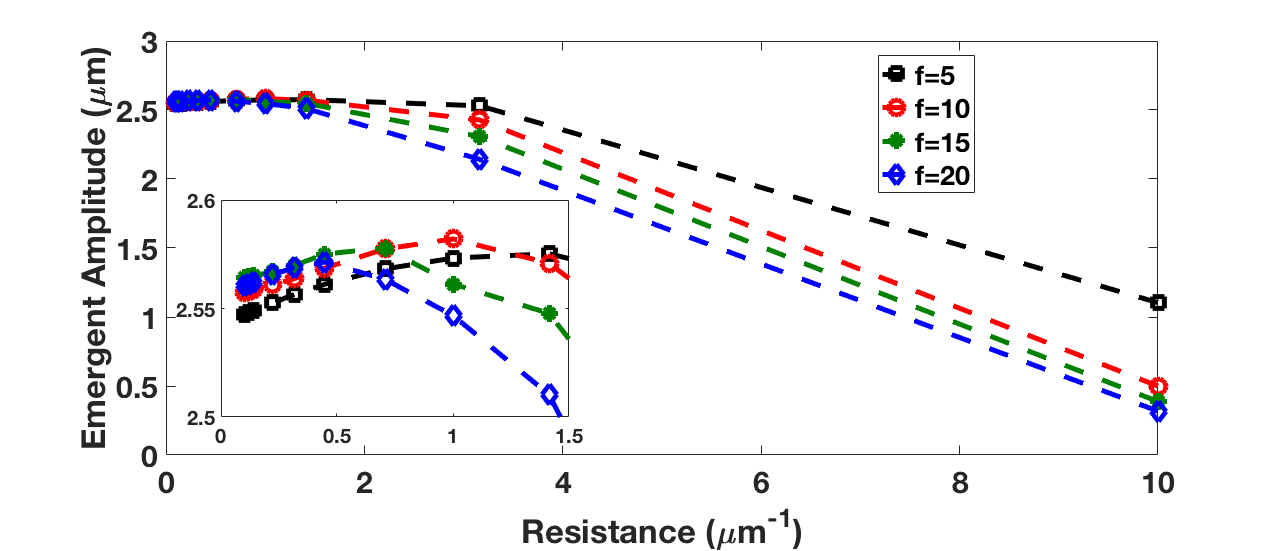}\\
\hspace{-5.2cm}{\bf (c)}\\ \vspace{-0.05cm}
\includegraphics[height=.3\linewidth,width=0.49\linewidth]{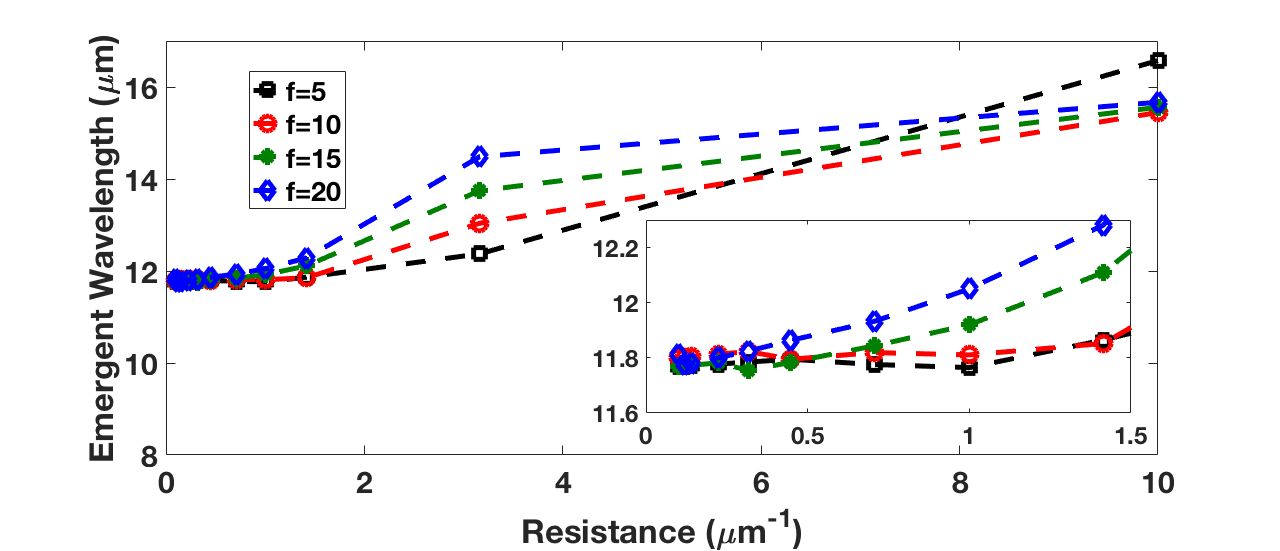}\\
\end{center}
\caption{Analyzing emergent behavior of a planar swimmer in a Brinkman fluid as a function of the resistance parameter $\alpha$: (a) swimming speeds, (b) amplitude, and (c) wavelength. The simulation results are denoted with symbols at $f=5, 10, 15,$ and $20$ (where $f=\sigma/2\pi$). Parameters used are: amplitude $b=4$, wavelength $\lambda=20$, regularization parameter $\varepsilon=5.454\triangle s$, and stiffness coefficients correspond to SIMS2 in Table \ref{tablecoefficientsWaveForms}. The insets on each graph highlight the behavior for small $\alpha$. }\label{EmergentParameterStudy}
\end{figure}
To further characterize the emergent behavior of swimmers propagating planar bending, we look at the emergent swimming speed, amplitude $b$, and wavelength $\lambda$ for swimmers with different beat frequencies $f=\sigma/2\pi$ in fluids with different resistance parameter $\alpha$.
First, we observe in figure \ref{EmergentParameterStudy}(a) that for all of the different beat frequencies, the maximum average swimming speed over the interval 0-1.2 s was obtained when $\alpha=1/\sqrt{0.5}\approx1.41$. For $\alpha<1$, the observed computational swimming speed decreases as $\alpha$ decreases. We also observe that the smaller the beat frequency is, the slower the swimmer, in agreement with the asymptotic swimming speed given in (\ref{asymp}). When $\alpha=10$, there is too much resistance in the fluid for the swimmers to achieve the preferred configurations. Figure \ref{EmergentParameterStudy}(b) shows that increased beat frequency results in a reduction of the achieved amplitude. In addition, swimmers are able to achieve a larger amplitude when $\alpha<1$. 
The emergent wavelength of the swimmers in a Brinkman fluid is also affected by $\alpha$ as shown in figure \ref{EmergentParameterStudy}(c). Here, the wavelength increases as resistance parameter $\alpha$ increases.  

\begin{figure}
 \centering
\includegraphics[height=.12\linewidth]{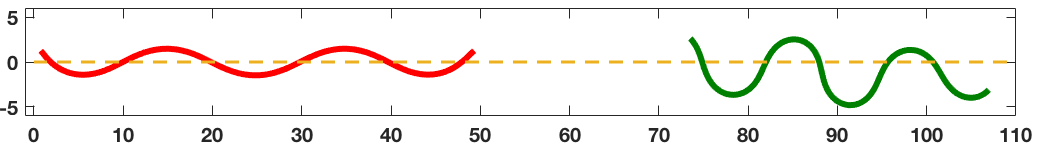}
\caption{\csr{A swimmer immersed in a Brinkman fluid with resistance $\alpha=1$ at $t=0.0012$ s on the left (red) and at $t=5$ s on the right (green). }}\label{OffCenterFigure}
\end{figure}

\csr{In figure \ref{OffCenterFigure}, a swimmer is shown at $t=0.0012$ s and $t=5$ s. The dashed line corresponds to $z=0$; we observe a downward tilt or yaw in the trajectory of the swimmer. The same phenomenon is observed at $t=5$ s regardless of the magnitude of the resistance; however, at higher resistance, the swimmers are not making as much forward progress. In addition, since we are propagating a preferred planar waveform, the average external torque component along the centerline is zero and the swimmer will remain in the plane.}

  \begin{figure}
 \centering
\includegraphics[height=.15\linewidth]{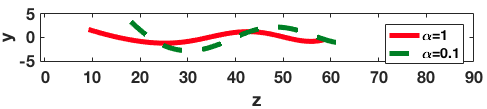}
\caption{Two separate simulations are shown at $t=0.6$ s for a swimmer in a Brinkman fluid with resistance $\alpha=1$ (solid line) and $\alpha=0.1$ (dotted line). The length of the rod is $L=50$ with wavelength $\lambda=40$, amplitude $b=4$, $f=20$, regularization parameter $\varepsilon=5.454\triangle s$, and stiffness parameters from SIMS 3 column of Table \ref{tablecoefficientsWaveForms}.}\label{SineWaveL40_SecondSetCoeff2}
\end{figure}

We also investigated the case of doubling the preferred wavelength to $\lambda=40$ while keeping the total length fixed at $L=50$ (representative of both human and sea urchin sperm). The wavelength of human sperm is, in general, less than that of sea urchin sperm, whereas the flagellum of a human sperm is stiffer than that of sea urchin \citep{lindemann2010flagellar,Smith09b,Woolley01}. With the same stiffness coefficients as in the previous case (SIMS2), we record little to no movement along the rod even in the case of small resistance. We speculate that the the stiffness of the swimmer may play a bigger role at larger wavelengths. To test this, we increase the stiffness coefficients and use the values reported in the SIMS3 column in Table \ref{tablecoefficientsWaveForms}, which are more representative of mammalian sperm. Representative results are shown in figure \ref{SineWaveL40_SecondSetCoeff2}. Here, we observe greater forward propulsion with smaller resistance. We also observe the figure-eight motions traced out by the endpoint  of the swimmer (not plotted). With the increased stiffness coefficients, trends for the emergent swimming speed, amplitude, and wavelength are similar to those shown in figure \ref{EmergentParameterStudy} as the resistance parameter $\alpha$ is varied. 

In summary, we have shown that for swimmers propagating planar waveforms, the emergent beatform may be different from the preferred one due to the resistance parameter $\alpha$, as well as the  rigidity of the swimmer. For smaller $\alpha$, the swimmer achieves a more obvious sinusoidal configuration and the emergent amplitude gets closer to the preferred one. If the rod is too stiff or if there is too much resistance in the fluid, the rod shows little to no forward motion, and the bending along its length is less likely to occur. 

\begin{figure}
\hspace{0.5cm}{\bf (a)} \hspace{18em}{\bf (b)}\\ \vspace{-0.3cm}
 \begin{center}
\includegraphics[width=.49\linewidth]{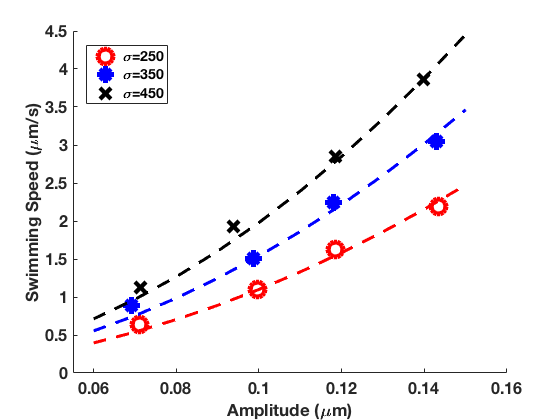}
\includegraphics[width=.49\linewidth]{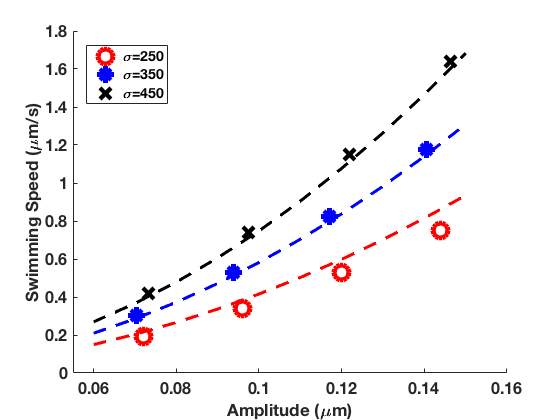}
\end{center}
\caption{Comparison of swimming speeds with resistance parameter $\alpha=1$ where symbols denote numerical results and the dashed lines are the asymptotic results. Swimming speeds are shown for helical bending in ({a}) and planar bending in ({b}). The length of the rod is $L=20$ ($N=301$ points) with wavelength $\lambda=20/3$, regularization parameter $\varepsilon=5.454\triangle s$, pitch $p=\lambda/2\pi$, and stiffness parameters from SIMS 2 column of Table \ref{tablecoefficientsWaveForms}.}\label{ComparePlanarSpiral}
\end{figure}

\subsection{Helical Bending}\label{HelixSmallAmp}
Similar to Section \ref{planarsection}, we first validate our method for the helical bending case by matching the numerical results to the asymptotic solutions for the case of helical bending with constant amplitude in both bending directions (referred to as spiral bending in \cite{Taylor52}). The rod is initialized as a right-handed helix parameterized as 
 \begin{eqnarray}
{\bf X}(s)=\left\{\frac{\kappa}{\kappa^2+\tau^2}\cos\left(\sqrt{\kappa^2+\tau^2}s\right), \frac{\kappa}{\kappa^2+\tau^2}\sin\left(\sqrt{\kappa^2+\tau^2}s\right), \frac{\tau}{\sqrt{\kappa^2+\tau^2}}s\right\},\label{hparameterization}
\end{eqnarray}
and the orthonormal triads are initialized as a rotation of the Frenet frame, given as
  \begin{eqnarray}
\left[
\begin{array}{ccccc}
{\bf D}^1(s)\\
{\bf D}^2(s)\\
{\bf D}^3(s)\\
\end{array}  \right]=
 \left[
\begin{array}{ccccc}
 \cos\phi & \sin\phi & 0 \\
-\sin\phi & \cos\phi & 0 \\
0& 0 &1\\
\end{array}  \right]
 \left[\begin{array}{ccccc}
{\bf N}\\
{\bf B}\\
{\bf T}\\
\end{array}  \right].\label{Frenet_DarbouxFrame}
\end{eqnarray}
Here, the normal vector is {\bf N}, the binormal vector is {\bf B}, the tangent vector is {\bf T}, and the angle of rotation is $\phi=-\tau s$ where $s$ is initialized as arc length. The detailed derivation of (\ref{hparameterization})--(\ref{Frenet_DarbouxFrame}) is in Appendix \ref{HelixDerived}.
The time-dependent preferred configuration in the form of a helix is given by
\begin{eqnarray*}
\Omega_1=\kappa\cos(\tau(s-Ut)), \hspace{2mm} \Omega_2=-\kappa\sin(\tau(s-Ut)), \hspace{2mm} \Omega_3=0,\label{hcurva}
 \end{eqnarray*} 
 where $U$ is the constant swimming velocity of the propagating wave and the beat frequency is $f=\sigma/2\pi=\tau U/2\pi$. The constants $\kappa$ and $\tau$ are the intrinsic curvature and torsion, respectively, depending on the radius $r_h$ and the pitch $p$ of the helix as follows: 
   \begin{eqnarray*}
r_h=\frac{\kappa}{\kappa^2+\tau^2}, \hspace{5mm}p=\frac{\tau}{\kappa^2+\tau^2},
\end{eqnarray*}
where $p=\lambda/2\pi$ for wavelength $\lambda$. This preferred curvature corresponds to helical bending propagating along the length of the swimmer, causing the swimmer to progress in the positive $z$-direction. Note that the curvature of the KR is $\Omega=\sqrt{\Omega_1^2+\Omega_2^2}=\kappa$.

\begin{figure}
 \centering
\includegraphics[height=.38\linewidth]{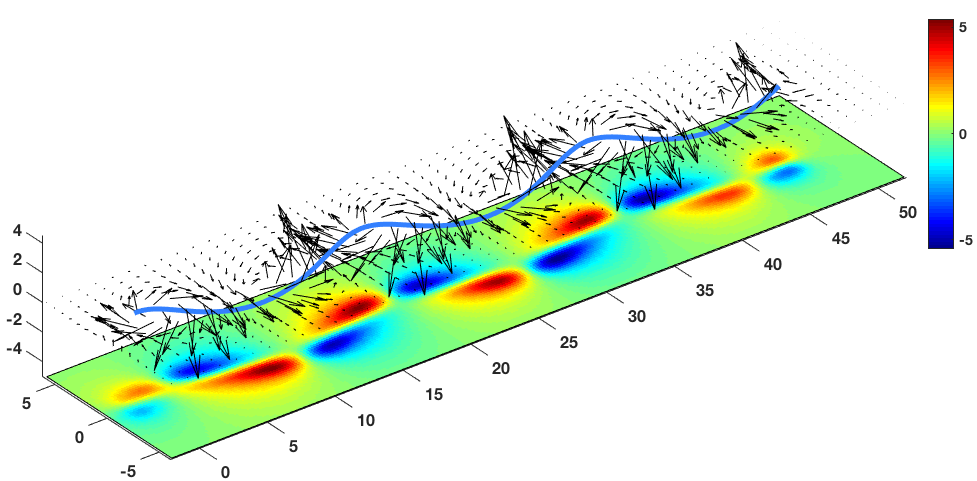}
\caption{The velocity field (black arrows) created by a swimmer in a Brinkman fluid with resistance parameter $\alpha=1$ at $t=0.5$ s. The corresponding pressure in the plane $x=0$ is shown below the swimmer with colorbar to the right (scaled by $p^*=10^4$). Parameters used are: $L=50$ ($N=301$ points), wavelength $\lambda=50/3$, regularization parameter $\varepsilon=5.454\triangle s$, $f=5$, $r_h=1$, and stiffness coefficients correspond to SIMS2 in Table  \ref{tablecoefficientsWaveForms}.}\label{PressVel_Helix}
\end{figure}

\begin{table}
\centering
\begin{tabular}{c|c|c|c|ccc}
\hline\hline 
Frequency&Results&$\alpha=1$ &  $\alpha=1/\sqrt{10}$& $\alpha=0.1$\\
\hline
$f=5$&Distance&0.9006&0.8701&0.8057\\
&Radius&0.5879&0.5506&0.5334\\
\hline
$f=20$&Distance&3.4491&3.3135&3.0728\\
&Radius&0.6327&0.5822&0.5584\\
\hline
$f=50$&Distance&7.9308&7.8252&7.2561\\
&Radius&0.6606&0.6010&0.5735\\
\hline
\end{tabular}
\caption{Distance traveled and emergent radii of helices in a Brinkman fluid with different resistance parameter $\alpha$ and frequency $f$ at $t=1.2$ s using $r_h=1$. Results for $f=20$ and $f=50$ correspond to figures \ref{HelixSigma20}--\ref{HelixSigma50}.}\label{EmergentHelixParameters}
\end{table}

In this test case, we explore swimming speeds for five different preferred radii ($r_h\ll 1$) and three different beat frequencies. 
In simulations, the numerical swimming speed is the average swimming speed in the $z$-direction and is plotted using the achieved radii at $t=0.6$ s. The computational swimming speeds are  compared to the previously derived asymptotic swimming speed for helical bending \citep{ho2016swimming}, which is twice the swimming speed of the planar bending case given in (\ref{asymp}) where we now replace $b$ with the radius $r_h$. 

The results are shown in figure \ref{ComparePlanarSpiral}(a) for $\alpha=1$, where we observe that the numerical swimming speeds (in marker points) match up well with the asymptotic swimming speeds (in dashed lines), increasing quadratically as the radius (or amplitude) increases. We observe similar swimming speed results when using a resistance parameter $\alpha$ in the range of $[0.01,10]$ (results not shown).  To investigate whether the emergent swimming speed of the helical swimmer is twice as fast as the corresponding planar swimmer (as predicted by the asymptotics), we use the same resistance parameter, beat frequency, and finite-length of the swimmer in the case of planar bending. The results are shown in figure \ref{ComparePlanarSpiral}(b) and indeed, the helical swimmer is faster than the planar swimmer. We observe that the speed up for finite-length helical vs. planar swimmers varies based on resistance parameter $\alpha$ and frequency $\sigma$. In figure \ref{ComparePlanarSpiral} for the case of $\alpha=1$, the helical bending swimmer is  3.1 times faster for $\sigma=250$ and 2.5 times faster for $\sigma=450$. 

\begin{figure}
\hspace{1.8cm}({\bf a})\hspace{10em}({\bf b})\hspace{10em}({\bf c})\\ \vspace{-0.3cm}
 \begin{center}
\includegraphics[width=.2\linewidth]{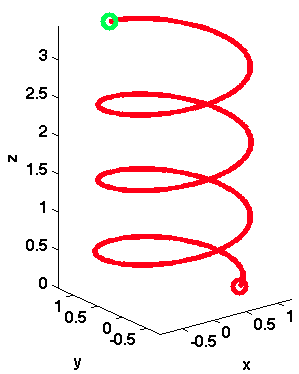}\hspace{3em}
\includegraphics[width=.2\linewidth]{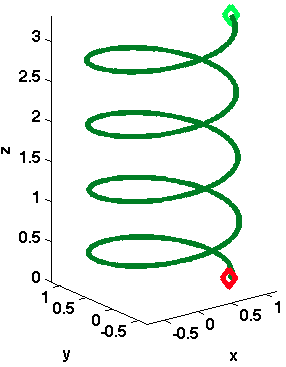}\hspace{3em}
\includegraphics[width=.2\linewidth]{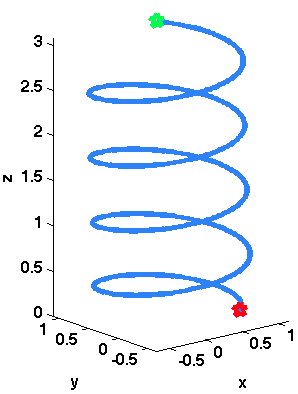}\\
\end{center}
\caption{The endpoint trajectories of a helix swimming in the positive $z$-direction in a Brinkman fluid with   ({a}) $\alpha=1$ (circle), ({b}) $\alpha=1/\sqrt{10}$ (diamond), and ({c}) $\alpha=0.1$ (star). Red and green markers correspond to the locations of the helix at $t=0$ and $t=1.2$ s, respectively. Parameters used are: $L=50$, radius $r_h=1$, wavelength $\lambda=50/3$, regularization parameter $\varepsilon=5.454\triangle s$, $f=20$, and stiffness coefficients correspond to SIMS2 in Table  \ref{tablecoefficientsWaveForms}.}\label{HelixSigma20}
\end{figure}

\begin{figure}
\hspace{1.8cm}({\bf a})\hspace{10em}({\bf b})\hspace{10em}({\bf c})\\ \vspace{-0.35cm}
 \begin{center}
\includegraphics[width=.2\linewidth]{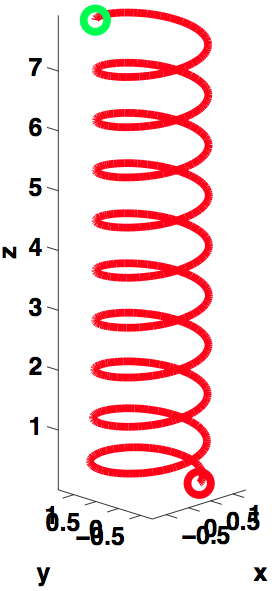}\hspace{3em}
\includegraphics[width=.2\linewidth]{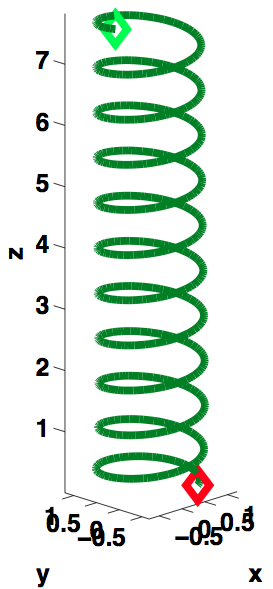}\hspace{3em}
\includegraphics[width=.2\linewidth]{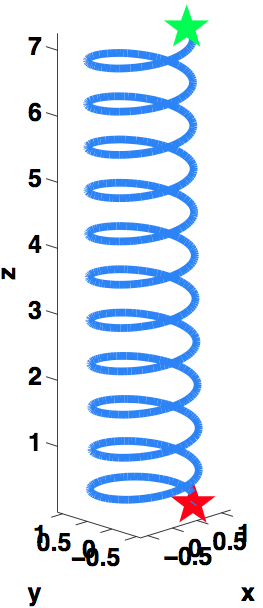}\\
\end{center}
\caption{The endpoint trajectories of a helix swimming in the positive $z$-direction in a Brinkman fluid with ({a}) $\alpha=1$ (circle), ({b}) $\alpha=1/\sqrt{10}$ (diamond), and ({c}) $\alpha=0.1$ (star). Red and green markers correspond to the locations of the helix at $t=0$ and $t=1.2$ s, respectively. Parameters used are: $L=50$, radius $r_h=1$, wavelength $\lambda=50/3$, regularization parameter $\varepsilon=5.454\triangle s$, $f=50$, and stiffness coefficients correspond to SIMS2 in Table  \ref{tablecoefficientsWaveForms}.}\label{HelixSigma50}
\end{figure}

\begin{figure}
\hspace{1cm}({\bf a}) \hspace{17em}({\bf b})
 \begin{center}
\includegraphics[width=.43\linewidth]{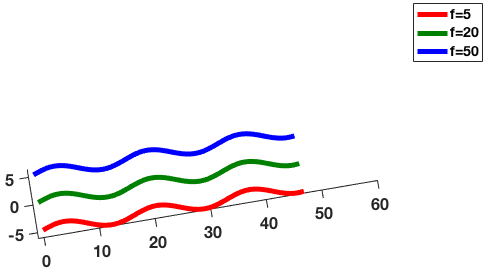}
\includegraphics[width=.43\linewidth]{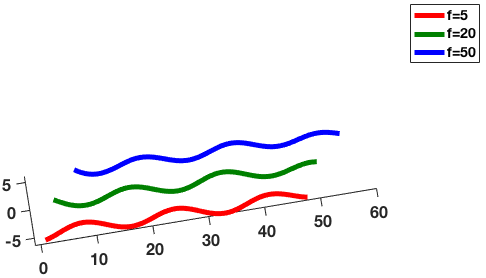}\\
 \end{center}
\caption{The side view showing bending along the length of helices with different beat frequency $f=5$ (red, bottom), $f=20$ (green, middle) and $f=50$ (blue, top) at (a) $t=0.0012$ s and (b) $t=1.2$ s. Parameters used are: $L=50$, radius $r_h=1$, wavelength $\lambda=50/3$, regularization parameter $\varepsilon=5.454\triangle s$, $\alpha=0.1$, and stiffness coefficients correspond to SIMS2 in Table  \ref{tablecoefficientsWaveForms}.}\label{HelixSideView}
\end{figure}

We now investigate emergent properties of a rod propagating helical bending as a function of the resistance parameter $\alpha$, using the same initialization as given in (\ref{hparameterization})--(\ref{hcurva}). 
In figure \ref{PressVel_Helix}, the flow field is shown around the swimmer with helical bending. Similar to the swimmer with planar bending (figure \ref{PressVel_Planar}), we observe local vortices of flow around the swimmer that change with the local concavity of the flagellar centerline. Trajectories for the end point (last point) of the helix propagating helical bending are shown in figures \ref{HelixSigma20}--\ref{HelixSigma50} and a side view of these helices is shown in figure \ref{HelixSideView}.  
In all cases, the swimmer traces out a helical trajectory which is consistent with the preferred helical bending. For a smaller beat frequency of $f=20$ in figure \ref{HelixSigma20}, the rod does not travel as far as with a higher frequency of $f=50$ in figure \ref{HelixSigma50}, as expected based on the asymptotic predictions. The actual distance traveled in 1.2 s is reported in Table \ref{EmergentHelixParameters}. In these simulations, the emergent radius of the swimmer is largest in the case of $\alpha=1$ (figure \ref{HelixSigma20}(a) and figure \ref{HelixSigma50}(a)) and smallest for the case of  $\alpha=0.1$  (figure \ref{HelixSigma20}(c) and figure \ref{HelixSigma50}(c)). Note that the preferred helix radius is $r_h=1$ but for this range of stiffness parameters in a Brinkman fluid, the swimmer is only able to achieve  a radius in the range of 0.5--0.6. 

\begin{figure}
  \hspace{0.85cm}{\bf (a)} \hspace{16.5em}{\bf (b)}\\ \vspace{-0.3cm}
  \begin{center}
\includegraphics[height=.3\linewidth,width=0.49\linewidth]{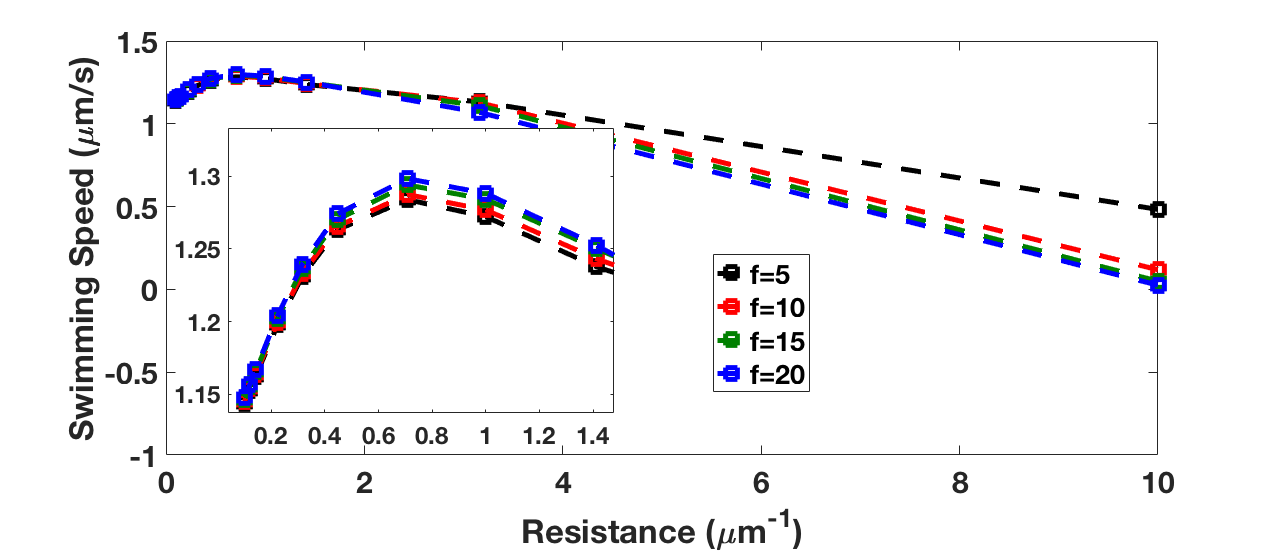}
\includegraphics[height=.3\linewidth,width=0.49\linewidth]{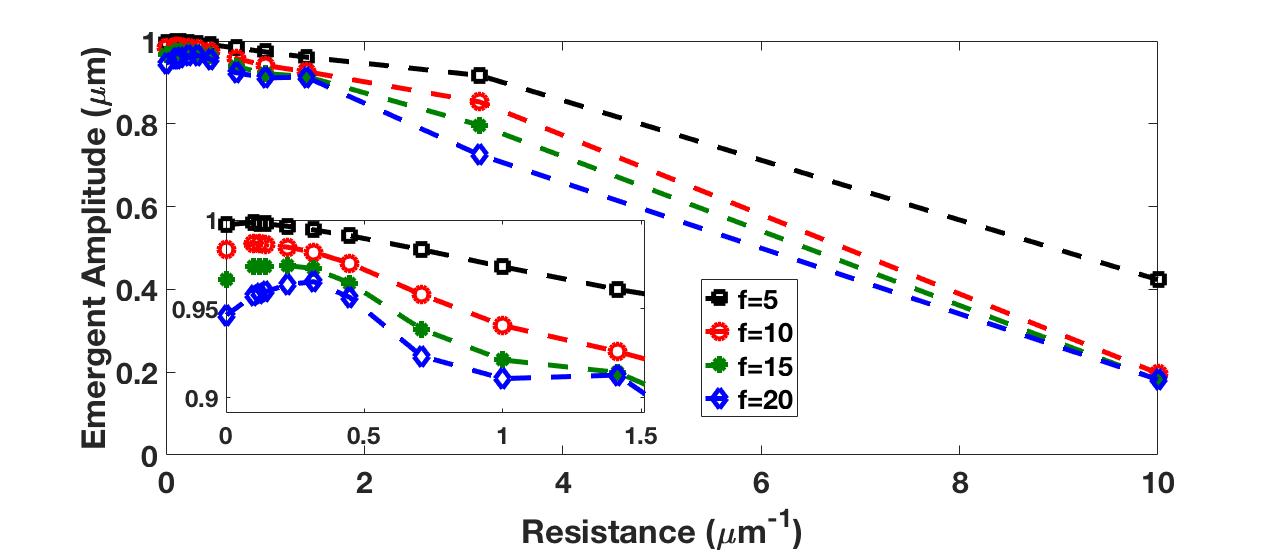}\\
\hspace{-5.5cm}{\bf (c)}\\
\includegraphics[height=.3\linewidth,width=0.49\linewidth]{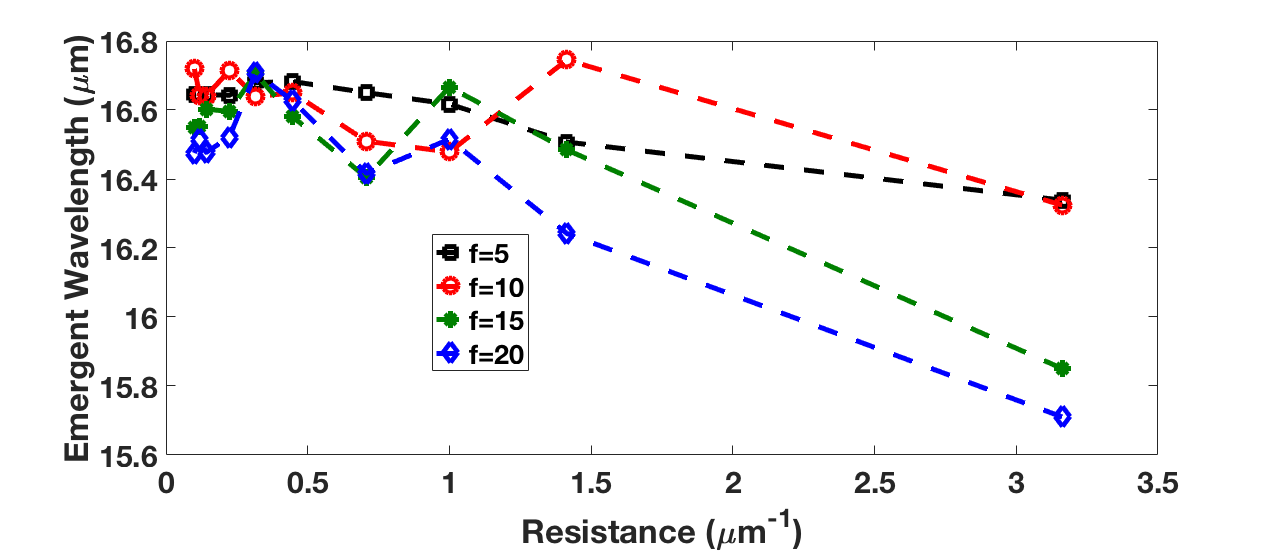}
\end{center}
\caption{\csr{Analyzing the emergent behavior of a helical swimmer in a Brinkman fluid as a function of resistance parameter $\alpha$: (a) swimming speeds, (b) amplitude, and (c) wavelength. The simulation results are denoted with symbols at $f=5, 10, 15,$ and $20$ (where $f=\sigma/2\pi$). Parameters used are: radius $r_h=1$, wavelength $\lambda=20$, regularization parameter $\varepsilon=5.454\triangle s$, and stiffness coefficients correspond to SIMS2 in Table \ref{tablecoefficientsWaveForms}. The insets on each graph highlight the behavior for small $\alpha$.} }\label{EmergentHelixStudy}
\end{figure}

\csr{To illustrate the effect of resistance parameter $\alpha$ on helical swimming, we calculate the achieved swimming speeds, radii, and wavelengths for a given preferred beatform for $\alpha$ in the range of 0.01--10. 
The Brinkman swimming speed results for different $\alpha$ in figure~\ref{EmergentHelixStudy}(a) are normalized by the Stokes case ($\alpha=0$). For small $\alpha$ in the range of 0.01--3, we observe that swimming speed is enhanced relative to the Stokes case. On the other hand, when the resistance parameter $\alpha$ is larger than 3, the helix swims slower than one in a fluid with no resistance. The maximum value of the swimming speed in this test case is at $\alpha=1/\sqrt{2}=0.707$. In figure~\ref{EmergentHelixStudy}(b), the emergent radii of the simulated helices are shown; the overall trend is that amplitude or radii decreases as the resistance increases. The emergent wavelength data shown in figure~\ref{EmergentHelixStudy}(c) fluctuates as the resistance changes. However, the fluctuations in wavelength are within $0.1-0.2$ $\mu$m, which could be due to post-processing to determine the wavelength. Thus, for resistance in the range of 0.01--1, we do not observe a significant change in the wavelength. For larger resistance, we do observe a small decrease in wavelength.  }
 
In summary, we have shown that for swimmers propagating helical bending, the emergent beatform may be different from the preferred one due to the resistance term in  the Brinkman fluid, which accounts for the presence of a sparse network of fibers. For smaller resistance parameter $\alpha$, the helical configuration has an emergent amplitude that is closer to the preferred one. In comparison to the planar bending case, the helical swimmer is faster for small amplitude, similar to the asymptotic predictions. For all $\alpha$, as the preferred amplitude or radius is increased, the helical swimmer becomes less able to achieve it and thus, swims slower.

\section{Discussion and Conclusions}
We have developed a numerical method to study 3D motility of a swimmer represented as a Kirchhoff rod (KR)  that is  immersed in a fluid governed by the incompressible Brinkman equations. The linear and angular velocity of the KR are derived using two different approaches, by regularizing fundamental solutions first or by choosing a regularization function and then determining the regularized fundamental solutions. This is an extension of previous work for the case of a KR immersed in a Stokesian fluid  \citep{Olson13}. 
At each point in time, we utilize the linearity of the Brinkman equation to write the solution as a superposition of fundamental solutions. However, since the forces are a function of the configuration of the KR, the coupled system and emergent swimming speeds are not linear with respect to resistance parameter $\alpha$. 

\csr{Through the choice of different time-dependent preferred curvature functions, the swimmer propagates planar or helical bending along the length of the rod. 
The actual force generation and bending of a sperm flagellum is  due to the local action of the dyneins \citep{lindemann2010flagellar}, which generate force independently and cause the microtubules to slide relative to one another. The preferred curvature that we utilize is a simplification that is motivated by experimental results of human sperm where the action of the dyneins results in a sinusoidal propagation of curvature \citep{Smith09b}. In our model, as the resistance parameter $\alpha$ is varied, we use the same preferred curvature but local force balance is altered by $\alpha$. Thus, for a given $\alpha$, we observe the \textit{emergent} flagellar waveform. Our model results are relevant for sperm motility in cases where varying the fluid properties does not change the ability of the dyneins to generate force and propagate a wave of curvature along the flagellum. The parameters for the preferred beat form were taken in the ranges reported for sperm and we vary the resistance parameter $\alpha$ in the range of 0 to 10 since it is well known that fiber radii and the volume fraction of the fibers can each vary by several orders of magnitude depending on the time in the menstrual cycle as well as the location in the mammalian reproductive tract \citep{Miki13,Smith09b}.

The stiffness coefficients in the model are chosen from a range of bending and shear moduli reported for sea urchin, rat, and bull sperm \citep{pelle2009mechanical,Lesich08,Schmitz-Lesich04}. The KR model is derived from force and torque balance on a cross section, which is simplified to a centerline representation when the rod length is much greater than the radius. Since we want to have a computational centerline that has the appropriate macroscopic moduli, we assign a shear and bending moduli based on experimental values, which is an approach that has been used before in other models of sperm and bacteria \citep{Lim04,Olson11a}. For bending amplitude and wavelength characteristic of human, we observe that swimming speeds are enhanced only when the stiffness of mammalian sperm is used (figure \ref{SineWaveL40_SecondSetCoeff2}), similar to previous computational results \citep{Olson15H}. When using the lower stiffness values of sea urchin sperm, we observe little to no forward progression (results not shown).  We hypothesize that in mammalian sperm, the additional accessory structures called the outer dense fibers \citep{gaffney2011mammalian}, are necessary to make the sperm flagellum stiffer and enhance mammalian sperm progression in fluids with a dense network of proteins.  

Although the asymptotic swimming speeds previously derived \citep{ho2016swimming} are able to capture the qualitative trends of swimming speed in terms of the dependence on resistance parameter $\alpha$ and amplitude, it often overestimates the actual swimming speed for shorter-length filaments and for filaments with larger preferred amplitudes. This is important to consider when using asymptotic swimming speeds to make predictions about the behavior of large amplitude, finite-length swimmers in a fluid with resistance. Our asymptotic analysis predicted that swimmers with helical bending are always two times faster than the corresponding planar swimmer. However, in our experience, this ratio is rarely observed and results are highlighted in Table \ref{planarspiralspeed} for $\alpha=1$. With the current computational method, we observe that helical swimmers with preferred amplitudes $b\ll1$, had swimming speeds greater than 2 times that of the corresponding planar swimmer and, in some cases, up to 3.3 times faster (Table \ref{planarspiralspeed}). In contrast, helical swimmers with larger preferred amplitudes achieve swimming speeds much less than the predicted ratio; in some cases almost 4 times less (results not shown). 
\begin{table}
\centering
\begin{tabular}{c|c|c|c|ccc}
\hline\hline
Frequency&Amplitude&Helical Bending & Planar  Bending & Ratio\\
& & Speed & Speed & \\
\hline
&0.075&0.6414&0.1923&3.34\\
$\sigma=250$&0.1&1.094&0.3399&3.21\\
&0.125&1.6226&0.5269&3.08\\
&0.15&2.1963&0.7516&2.92\\
 \hline
&0.075&0.8862&0.3002&2.95\\
$\sigma=350$&0.1&1.5122&0.5307&2.85\\
&0.125&2.2443&0.8231&2.72\\
&0.15&3.0403&1.1747&2.59\\
\hline
&0.075&1.1253&0.4188&2.68\\
$\sigma=450$&0.1&1.9214&0.7404&2.59\\
&0.125&2.8534&1.1487&2.48\\
&0.15&3.8676&1.6397&2.36\\
\hline
\end{tabular}
\caption{\textit{The average swimming speeds for helical and planar bending are shown for various preferred frequency $f=\sigma/2\pi$ and amplitude $b$ for the case of $\alpha=1$. The ratio between the swimming speed of helical and planar bending is also calculated}.}\label{planarspiralspeed}
\end{table}

In the case of planar and helical bending, we observe an emergent amplitude that generally decreases as the resistance parameter $\alpha$ increases, even with the same preferred configuration (figure \ref{EmergentParameterStudy}(A) and \ref{EmergentHelixStudy}(A)). Due to this decreased amplitude, a decrease in swimming speed for $\alpha>2$ was also observed. This is in contrast to  the previously obtained asymptotic results where bending kinematics (amplitude) were prescribed \citep{ho2016swimming} and it was possible to obtain higher swimming speeds in fluids with more fibers (larger $\alpha$). \csr{In this previous analysis, it was found that the work required to generate the prescribed amplitude increased greatly as $\alpha$ or prescribed amplitude increased.} However, in reality, it may not be possible to achieve a swimmer with higher amplitude because microorganisms are not able to generate that much work to maintain  the prescribed bending at such high resistance. To further understand the relation between the numerical results presented here and the previous asymptotic analysis, we investigate the asymptotic swimming speeds for a fixed value of work.  Equation (\ref{asymp}) can be rewritten as
$$\frac{U_{\infty}}{U}=\frac{1}{2}W_{\infty}\left[K_0(\zeta_1)-\frac{1}{2}\left(\frac{k^2}{\alpha^2}+1\right)\log\left(1+\frac{\alpha^2}{k^2}\right)\right],$$
in terms of the nondimensional work $W_{\infty}=\overline{W}/\mu\pi U^2$, where $\overline{W}$ is detailed in \citet{ho2016swimming}. We fix $\zeta_1=kr_a=0.03$ and consider three different fixed values of $W_{\infty}$ as 0.15, 0.2, 0.25. Figure \ref{thrustwork}(a) shows the relation between the nondimensional swimming speed $U_{\infty}/U$ and the scaled resistance parameter $\alpha/k$. 
We observe that for a larger fixed value of work, there is a faster swimming speed.  This means that if a swimmer can achieve a higher amount of work, it would be able to swim faster. On the other hand, as the scaled resistance increases, the swimming speed decreases for all three cases with a fixed value of work. This matches our numerical results where swimmers in fluids with large resistance swim slower; this is due to the fact that the forces are based on an energy formulation where emergent beatforms minimize energy in the system \citep{Lim08}. Hence, work is not able to significantly increase in our 3D Brinkman simulations with the KR, causing amplitude and swimming speed to decrease at high $\alpha$.   

 \begin{figure}
\hspace{0.5cm}{\bf(a)}\hspace{4.85cm}{\bf(b)}\\ \vspace{-.3cm}
\begin{center}
\includegraphics[width=0.48\textwidth]{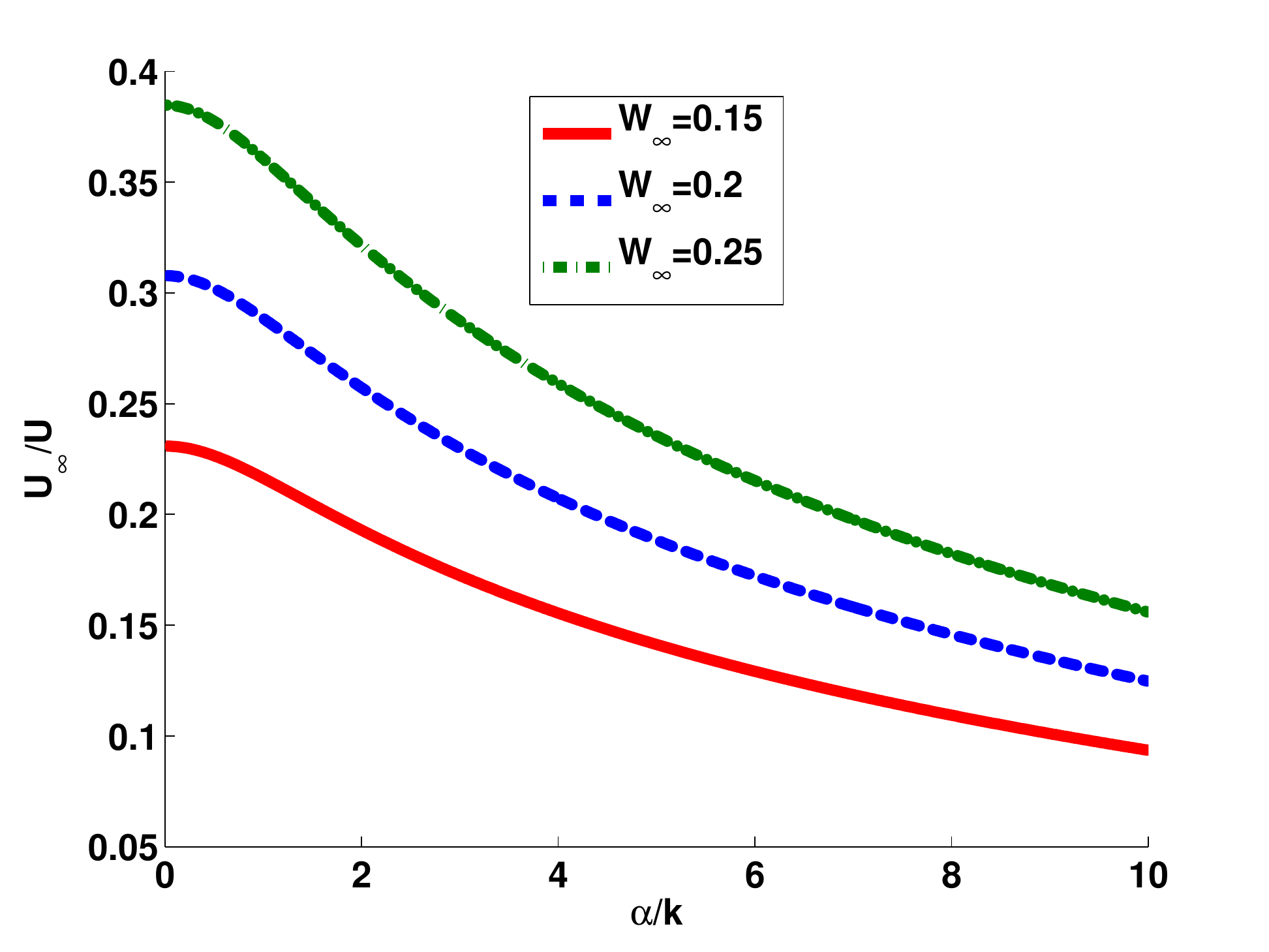}
\includegraphics[width=0.48\textwidth]{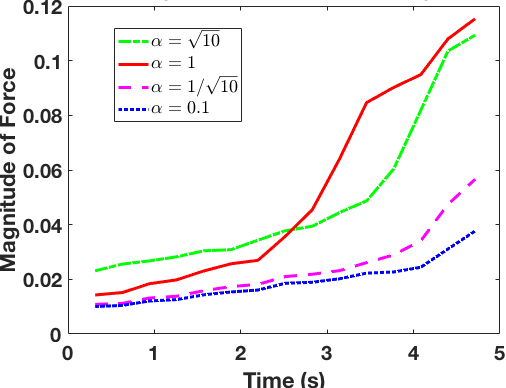}\\
\hspace{-6cm}{\bf{(c)}}\\
\includegraphics[width=0.52\textwidth]{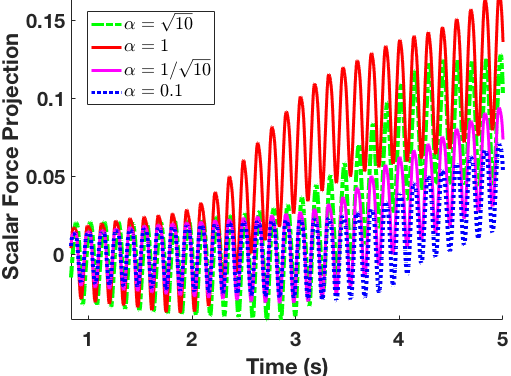}
\end{center}
\caption{(a) The nondimensional swimming speed $U_{\infty}/U$ in terms of the scaled resistance $\alpha/k$ for fixed values of work $W_{\infty}$ are plotted for $\zeta_1=0.03$. For four values of the resistance parameter $\alpha$ up to $t=5$ s, we show metrics for propulsive thrust: (b) magnitude of forces along the flagella in the direction of swimming averaged over a beat period and (c) scalar projection of forces in the direction of swimming at the tail end of the swimmer. The parameters used for (b)-(c) correspond to those used in figure \ref{SnapShot1}.}
\label{thrustwork}
\end{figure}  

Experimental studies have shown that emergent waveforms and swimming speeds vary greatly depending on the fluid environment \citep{Smith09b,Suarez92}. Since it is both time intensive and costly to perform many experiments tracking sperm motility, the computational framework we present can be used to systematically study 3D flagellar waveforms as protein volume fraction $\varphi$ or resistance parameter $\alpha$ is varied. In this study, we investigate the emergent waveforms and swimming speeds in the cases of planar and helical bending. We  observe non-monotonic swimming speeds with respect to the resistance parameter $\alpha$ (figure \ref{EmergentParameterStudy}(A) and \ref{EmergentHelixStudy}(A)). 
The enhancement in swimming for a range of $\alpha$ may be explained as follows. In the fluid with low resistance (when $\alpha\to0$, this corresponds to no fibers or Stokes flow), the swimmer is able to bend freely (without any effects from the obstacles) and achieve the preferred bending. When the volume fraction of fibers in the fluid is small, the fibers alter or modify the flow in a way that gives an extra push to enhance the swimming velocity. As the resistance parameter $\alpha$ increases, the flow decays in a smaller region as described by the screening length, given by $\sqrt{\gamma}=1/{\alpha}$. This then results in a different local force and torque balance along the flagellum, with a decrease in the observed amplitude for larger $\alpha$ (figure \ref{SnapShot1}). 

In order to further characterize this swimming enhancement for small resistance parameter $\alpha$, we will look at an indicator of thrust or propulsive forces since there is not an easy and direct calculation of propulsive forces in a model of force-free swimming with fully coupled and emergent flagellar waveforms. For the planar case, the propulsive forces we report are the body forces of the swimmer from (\ref{thrustforce}) in the direction of swimming. This is calculated as a projection of the forces onto the centerline (line connecting the center of mass and first point or head of swimmer), calculated for each time point. To understand how forces vary in time, in figure \ref{thrustwork}(b) we look at the integral of the average magnitude of the forces along the length of the flagella over a beat period. For the planar case, recall that the swimmer is initialized as a straight rod and then deforms and interacts with the fluid due to the local force and torque balance to achieve the preferred beat form. As can be seen in figure \ref{SnapShot1}(a)-(b) and \ref{thrustwork}(b), at $t=$0.0012 and 0.12 s, there is not much of a difference in swimming speeds and magnitude of forces for the different cases of $\alpha$ shown. For later times, at $t=$0.6 and 1.2 s, the swimmers have achieved larger amplitude waveforms and the $\alpha=1$ and $\alpha=\sqrt{10}$ have progressed further due to the increased force magnitude in the direction of swimming. Since a sperm flagellum can be considered a pusher, the swimming propulsion or thrust will be from the rear or tail end of the swimmer \citep{Pak}. To further characterize thrust or propulsive forces, we also calculate the scalar projection of the force onto the centerline at the end of the swimmer as shown in figure \ref{thrustwork}(c). Here, the forces oscillate in time due to the propagating curvature. Looking at $\alpha=1$ and $\sqrt{10}$, we observe the largest thrust in the swimming direction when $\alpha=1$ and the smallest thrust in the case of $\alpha=0.1$. There is an upward trend in the propulsive forces, similar to the average over a beat period shown in figure \ref{thrustwork}(b). We observe similar trends for a range of parameters for both the planar and helical beat form. Another way to understand these emergent swimming speeds and waveforms is to consider the damping stress, the term $-\mu\alpha^2{\bf u}$ in (\ref{BrEq}), which dominates in comparison to the viscous shear stress effects as $\alpha$ is further increased \citep{ingham2005transport,kaviany2012principles}. Thus, for  large $\alpha$ ($\alpha>2$), the shear stress is only generated over a very short distance \citep{kaviany2012principles}, preventing filaments from reaching their preferred configuration and decreasing propulsive forces in the direction of swimming. 

We note that this non-monotonic trend in swimming speeds has also been observed for a swimmer propagating planar bending in both 2D and 3D Brinkman fluids \citep{Cortez10,Olson15H,leiderman2016swimming}. However, in those studies the bending was only allowed in the plane and we know from previous computational analyses that there can be differences in emergent behavior when comparing 2D and 3D models \citep{Olson15}. Since this non-monotonic behavior has now been observed in a similar range of $\alpha$ in both 2D and 3D with different force models, we believe that this is a robust phenomenon. Additionally, since we are fully accounting for the 3D nature of the flagellar beatform in this Kirchhoff rod modeling framework, we believe that the range of $\alpha$ corresponding to enhancements in swimming speed, $\alpha\in(0,2)$, is more representative of the relevant range. Using this range of $\alpha$, we can back out different possible combinations of fiber volume fraction $\varphi$ and fiber radii $a_f$, and then determine the interfiber spacing $D$ using (\ref{fibrouseq})--(\ref{ratiointerfiber}), which are illustrated in figure \ref{EstimateGraphPic}(a) and (b).  For example, $\alpha=1$ ($\gamma=1$) could be realized with  a fiber radii $a_f=0.2$ microns, which would give a volume fraction of $\varphi\sim$0.021 and an estimated interfiber spacing of $D\sim5$ microns. Similarly, $\alpha=2$ ($\gamma=1/4$) could be realized with fiber radii $a_f=0.1$ microns, which would give a volume fraction of $\varphi\sim0.0025$ and an estimated interfiber spacing of $D\sim10$ microns. We note that for the development of artificial microswimmers, there may be applications where fluids could be created with stationary fibers of radii on the range of 0.3--2 microns, which would give an interfiber spacing of $D=10-200$ microns. Vaginal fluid fiber radii have been reported as large as 0.35 microns \citep{Rutllant05}, which would correspond to an interfiber spacing $D\sim40$ microns. At the larger end of this scale, a human sperm of length 50 microns and wavelength of 10-50 microns \citep{Smith09b}, would be able to swim freely through the stationary network with little to no interactions with the fibers. If fiber radii were an order of magnitude smaller, this would put interfiber spacing on the order of a micron. The width of the flagellum would fit easily between the fibers but the assumption that the flagellum is not directly interacting with the fibers (either sliding along fibers or hitting fibers) would need to be accounted for if interfiber spacing was actually on the order of a micron.  We note that the interfiber spacing $D$ in (\ref{ratiointerfiber}) is an \textit{estimate} whose derivation is based on the assumption of randomly oriented fibers. Recent experimental results have shown that human cervicovaginal mucus actually has pore spacing that is much larger than that predicted by using mucin fiber diameter and volume fraction when assuming randomly oriented fibers \citep{Lai09}. The larger pore spacing observed was around 18 microns, a range where sperm interactions with the fibers could still be assumed minimal. Thus, additional analysis to determine a better approximation for interfiber spacing for different arrangements of proteins as well as additional experiments to determine a more realistic average fiber spacing of cervical and vaginal  fluid are necessary to determine whether the enhancement observed for $\alpha\sim1$ is in a biologically relevant regime. 

Experiments have recorded emergent flagellar waveforms of human sperm in  a low viscosity medium consisting of a salt solution with 0.45\% serum albumin (very small volume fraction of albumin protein fibers in the fluid) and a high viscosity medium that was obtained by adding 1\% methylcellulose (MC) to the low viscosity solution (corresponding to the addition of polymer chains in the fluid) \citep{Smith09b}. In the Brinkman equation, we are assuming that the fluid viscosity is remaining constant (i.e., the dynamic and effective viscosity are equal) and account directly for the volume fraction of proteins through the resistance parameter $\alpha$ \citep{Auriault09}. Thus, as a rough approximation, the experiments where viscosity of the gel is increased correlate to our computational results where $\alpha$, and hence volume fraction of proteins or polymers, is increased.  In the case of planar and helical swimmers, as shown in figures \ref{EmergentParameterStudy} and \ref{EmergentHelixStudy}, we observe a general trend that amplitude decreases  as $\alpha$ increases. This trend is similar to results of human sperm swimming in MC gels where the higher viscosity gel had a decreased amplitude \citep{Smith09b}. In contrast, this experimental study showed a significant decrease in the wavelength with the higher viscosity MC gels. In our simulations, we observe an increase in wavelength for the planar swimmer (figure \ref{EmergentParameterStudy}) and a very small decrease in wavelength for the helical swimmer (figure \ref{EmergentHelixStudy}). In our computational model, as the resistance parameter $\alpha$ is varied, this will cause the local force balance to be altered and as a result, we observe the emergent wavelength. The differences  in emergent flagellar wavelength between experiments and these computational results could be due to several factors. The higher viscosity MC gels also have a small amount of elasticity (relaxation time of 0.006 s and 0.017 s \citep{Smith09b}) and the effective and dynamic viscosity are most likely not equal, which we do not account for in our model. }

Sperm cells have been observed to `roll' as they swim (simultaneous rotation of the sperm cell body and flagellum) \citep{Babcock14,Smith09b}. Specifically, human sperm were found to increase rolling from 1.5 Hz to 10 Hz and decrease amplitude as the viscosity of methylcellulose solutions was decreased \citep{Smith09b}. \csr{The tilt or yaw of human sperm was also found to decrease in a gel with higher viscosity \citep{Smith09b}. Similar to previous computational studies \citep{olson2011coupling,Smith09Surf}, we observe the swimmer does not swim exactly straight, but does tilt downward as shown in figure \ref{OffCenterFigure}. We note that for the planar swimmer, in our computational model, the average external torque component along the centerline is zero and the swimmer will remain in the plane. The angular velocity shows that the KR is rotating, but at a rate much smaller than the experimental data. In addition, simulations with a prescribed beatform swimming in a Brinkman fluid showed an increased angular velocity with increased prescribed amplitude \citep{ho2016swimming}. However, in our model presented here, we do not observe an increase in angular velocity. With an emergent beatform, the swimmer does not have as much angular velocity since the local force balance allows the swimmer to achieve a different preferred amplitude and does not have to rotate as much to achieve a prescribed amplitude. Our results do differ from experiments in terms of the rolling rate. This could be due to the fact that we are not accounting for the dynamics of a cell body.} 
A computational study of 3D finite-length swimmers with cell bodies and emergent kinematics is necessary to fully understand swimming speed and angular velocity as a function of the resistance. This will be a focus of future studies.

Lastly, the velocity field in this regularized method is given as the superposition of the flow velocity generated by $N$ forces exerted on the fluid by a swimmer whose centerline is represented as a KR. As described in Appendix \ref{limits}, as the resistance parameter $\alpha\to0$, we obtain the solution corresponding to that of Stokes equations. We note that the velocity in Eq.~(\ref{uvel_B})--(\ref{wvel_B}) is an approximation to the integral equation using a quadrature rule. Thus, in addition to a regularization error, the evaluation of the integral also introduces a quadrature error \citep{Cortez05,Cortez10}. A recent study of flow past a cylinder (2D) using the method of regularized Brinkmanlets showed the error of the velocity field depends on both the regularization parameter $\varepsilon$ and the resistance $\alpha$ \citep{leiderman2016swimming}. In fact, the minimum computed error increased as $\alpha$ increased. It would be interesting to further investigate the errors (discretization error and quadrature error) on and off a 3D KR  to understand how different blob functions, regularization parameters, and resistance $\alpha$ affect the errors.

\section*{Acknowledgements}
The work of N. Ho was funded, in part, by National Science Foundation grant DMS-1413110. S.D. Olson, was funded, in part, by National Science Foundation grants DMS-1413110 and 1455270. K. Leiderman was funded, in part, by National Science Foundation grant DMS-1413078. The authors thank R. Cortez for helpful discussions.

\appendix
\section{Linear \& Angular Velocity}\label{CoefficientsFunctions}
 The details for the derivation of the linear and angular velocity are given here and will depend on the choice of the blob function $\phi_{\varepsilon}(r)$ or the regularized functions $B_{\varepsilon}(r)$ and $G_{\varepsilon}(r)$ where $\Delta G_{\varepsilon}(r)=\phi_{\varepsilon}(r)$ and $(\Delta-\alpha^2)B_{\varepsilon}(r)=G_{\varepsilon}(r)$. Consider the linear velocity in (\ref{linearvel}) for a constant point force ${\bf f}_c$ and torque ${\bf m}_c$, both applied at ${\bf X}_c$ where $\hat{\mathbf{x}}={\bf x}-{\bf X}_c$ and $r=||\hat{\bf x}||$.
The gradient and laplacian terms on the right hand side of  (\ref{linearvel}) can be expanded as 
    \begin{eqnarray*}
 \mu{\bf u}({\bf x})&=&({\bf f}_c\cdot{\bf \hat{x}}){\bf \hat{x}}\frac{rB''_{\varepsilon}(r)-B'_{\varepsilon}(r)}{r^3}+{\bf f}_c\frac{B'_{\varepsilon}(r)}{r}-{\bf f}_c\left(B''_{\varepsilon}(r)+\frac{2}{r}B_{\varepsilon}(r)\right)\\
 &&\hspace{2in}-\frac{1}{2}\alpha^2B'_{\varepsilon}(r)\frac{{\bf \hat{x}}}{r}\times {\bf m}_c-\frac{1}{2}G'_{\varepsilon}(r)\frac{{\bf \hat{x}}}{r}\times {\bf m}_c,\\
 &=&({\bf f}_c\cdot{\bf \hat{x}}){\bf \hat{x}}\frac{rB''_{\varepsilon}(r)-B'_{\varepsilon}(r)}{r^3}+{\bf f}_c\left(-B''_{\varepsilon}(r)-\frac{B_{\varepsilon}(r)}{r}\right)\\
  &&\hspace{2in}+\frac{1}{2}\alpha^2({\bf m}_c\times{\bf \hat{x}})\frac{B'_{\varepsilon}(r)}{r}+\frac{1}{2}({\bf m}_c\times{\bf \hat{x}})\frac{G'_{\varepsilon}(r)}{r}.
  \end{eqnarray*} 
Then, using (\ref{hgeneral})--(\ref{dgeneral}), the linear velocity can be rewritten as given in (\ref{uvel_B}).
Following the same derivation, the angular velocity from (\ref{angularvel}) can be expanded as
  \begin{eqnarray*}
  \mu{\boldsymbol \omega}({\bf x})  &=&\frac{1}{2}\alpha^2({\bf f}_c\times{\bf \hat{x}})\frac{B'_{\varepsilon}(r)}{r}+\frac{1}{2}({\bf f}_c\times{\bf \hat{x}})\frac{G'_{\varepsilon}(r)}{r}-\frac{1}{4}\alpha^2({\bf m}_c\cdot{\bf \hat{x}}){\bf \hat{x}}\frac{rB''_{\varepsilon}(r)-B'_{\varepsilon}(r)}{r^3}\\
    &&\hspace{10mm}+\frac{1}{4}\alpha^2\left(B''_{\varepsilon}(r)+\frac{B_{\varepsilon}(r)}{r}\right){\bf m}_c-\frac{1}{4}({\bf m}_c\cdot{\bf \hat{x}}){\bf \hat{x}}\frac{rG''_{\varepsilon}(r)-G'_{\varepsilon}(r)}{r^3}+\frac{1}{4}{\bf m}_c\left(\Delta G_{\varepsilon}-\frac{G'_{\varepsilon}(r)}{r}\right),\\
      &=&\frac{1}{2}({\bf f}_c\times{\bf \hat{x}})\left[\alpha^2\frac{B'_{\varepsilon}(r)}{r}+\frac{G'_{\varepsilon}(r)}{r}\right]-\frac{1}{4}\alpha^2\left[({\bf m}_c\cdot{\bf \hat{x}}){\bf \hat{x}}\frac{rB''_{\varepsilon}(r)-B'_{\varepsilon}(r)}{r^3}-\left(B''_{\varepsilon}(r)+\frac{B_{\varepsilon}(r)}{r}\right){\bf m}_c\right]\\
    &&\hspace{15mm}+\frac{1}{4}\left[\left(\phi_{\varepsilon}-\frac{G'_{\varepsilon}(r)}{r}\right){\bf m}_c-({\bf m}_c\cdot{\bf \hat{x}}){\bf \hat{x}}\frac{rG''_{\varepsilon}(r)-G'_{\varepsilon}(r)}{r^3}\right].
\end{eqnarray*} 
 Then, using (\ref{hgeneral})--(\ref{dgeneral}), we arrive at (\ref{wvel_B}).

  \subsection{Option 1: Regularizing the Fundamental Solutions}\label{option1app}
  The regularized fundamental solutions $G_{\varepsilon}$ and $B_{\varepsilon}$ are given in (\ref{fundreg}) and we arrive at the following coefficient equations when using (\ref{hgeneral})--(\ref{dgeneral}):
     \begin{eqnarray}
H_1^{\varepsilon}(r)&=&\frac{e^{-\alpha R}}{4\pi R}\left(\frac{1}{\alpha^2R^2}+\frac{1}{\alpha R}+1\right)-\frac{1}{4\pi\alpha^2R^3},\label{H1singular}\\
H_2^{\varepsilon}(r)&=&-\frac{e^{-\alpha R}}{4\pi R^3}\left(\frac{1}{\alpha^2R^2}+\frac{1}{\alpha R}+1\right)+\frac{3}{4\pi\alpha^2R^5},\label{H2singular}\\
Q_2^{\varepsilon}(r)&=&\frac{e^{-\alpha R}}{4\pi \alpha^2R^3}\left(1+\alpha R\right)-\frac{1}{4\pi \alpha^2R^3},\\
Q_1^{\varepsilon}(r)&=&\frac{5\varepsilon^2}{R^2}H_2^{\varepsilon}(r)-\frac{\alpha^2\varepsilon^2}{R^2}Q_2^{\varepsilon}(r)+\frac{1}{4\pi R^3}\left(1-\frac{\varepsilon^2}{R^2}\right),\\
D_2^{\varepsilon}(r)&=&\frac{35\varepsilon^2}{R^4}H_2^{\varepsilon}(r)-\frac{10\alpha^2\varepsilon^2}{R^4}Q_2^{\varepsilon}(r)-\frac{1}{4\pi R^5}\left(\frac{10\varepsilon^2}{R^2}+\alpha^2\varepsilon^2e^{\alpha R}-3\right),\label{D2reg}
 \end{eqnarray}
 and $D_1^{\varepsilon}(r)=\psi_{\varepsilon}-Q_1^{\varepsilon}$ for 
 corresponding blob function $\psi_{\varepsilon}$. Using $\Delta G_{\varepsilon}=\psi_{\varepsilon}$, the blob function is then
  \begin{eqnarray}
\psi_{\varepsilon}(r)&=&\frac{\varepsilon^2}{4\pi \alpha^2R^9}\left\{3(\alpha^2R^4+35\varepsilon^4-20\varepsilon^2R^2)-e^{-\alpha R}\left[\alpha^2R^2(1+\alpha R)(\alpha\varepsilon^2R+9\varepsilon^2-7R^2)\right.\right.\nonumber\\
&&\left.\left.\hspace{30mm}+5(7\varepsilon^2-4R^2)(\alpha^2R^2+3+3\alpha R)-30\alpha^2\varepsilon^2R^2\right]\right\}.\label{BloBFunction1}
 \end{eqnarray}

 \subsection{Option 2: Choosing a blob function}\label{option2app}
In (\ref{gblob})--(\ref{bblob}), for the blob function specified in (\ref{NganBlobFunction}), we can use (\ref{hgeneral})--(\ref{dgeneral}) to determine $H_i^{\varepsilon},~ Q_i^{\varepsilon},~ D_i^{\varepsilon}$  for $i=1,2$. These coefficients functions are given as
     \begin{eqnarray}
H_1^{\varepsilon}(r)&=& 
\left\{
    \begin{array}{ll}
\displaystyle \frac{(\varepsilon^2+2r^2)e^{-r^2/\varepsilon^2}}{2\pi^{3/2}\alpha^2\varepsilon^3r^2}\left(1-e^{-\alpha^2\varepsilon^2/4}\right)+\frac{1+\alpha r+\alpha^2r^2}{8\pi\alpha^2r^3}\textrm{erfc}\left(\frac{\alpha\varepsilon}{2}-\frac{r}{\varepsilon}\right)e^{-\alpha r}\\\\
\displaystyle\hspace{5mm}-\frac{1}{4\pi\alpha^2r^3}\left[1-\textrm{erfc}\left(\frac{r}{\varepsilon}\right)\right]-\frac{1-\alpha r+\alpha^2r^2}{8\pi\alpha^2r^3}\textrm{erfc}\left(\frac{r}{\varepsilon}+\frac{\alpha\varepsilon}{2}\right)e^{\alpha r}, \hfill r> 0 \\\\\label{H1blob}
\displaystyle\frac{2}{3\pi^{3/2}\alpha^2\varepsilon^3}\left(1-e^{-\alpha^2\varepsilon^2/4}\right)-\frac{\alpha}{6\pi}\left[1-\textrm{erf}\left(\frac{\alpha\varepsilon}{2}\right)\right]+\frac{1}{3\pi^{3/2}\varepsilon}e^{-\alpha^2\varepsilon^2/4},\hfill r=0
    \end{array} \right.\\\nonumber\\
H_2^{\varepsilon}(r)&=&
\left\{    \begin{array}{ll}
\displaystyle-\frac{\left(3\varepsilon^2+2r^2\right)e^{-r^2/\varepsilon^2}}{2\pi^{3/2}\alpha^2\varepsilon^3r^4}\left(1-e^{-\alpha^2\varepsilon^2/4}\right)-\frac{3+3\alpha r+\alpha^2r^2}{8\pi\alpha^2r^5}\textrm{erfc}\left(\frac{\alpha\varepsilon}{2}-\frac{r}{\varepsilon}\right)e^{-\alpha r}\\\\
\displaystyle\hspace{10mm}+\frac{3}{4\pi\alpha^2r^5}\textrm{erf}\left(\frac{r}{\varepsilon}\right)+\frac{3-3\alpha r+\alpha^2r^2}{8\pi\alpha^2r^5}\textrm{erfc}\left(\frac{\alpha\varepsilon}{2}+\frac{r}{\varepsilon}\right)e^{\alpha r},\hfill r> 0 \\\\
\displaystyle \frac{e^{-\alpha^2\varepsilon^2/4}}{30\pi^{3/2}\varepsilon^3}\left(2-\alpha^2\varepsilon^2\right)+\frac{2}{5\pi^{3/2}\alpha^2\varepsilon^5}\left(1-e^{-\alpha^2\varepsilon^2/4}\right)\\
\displaystyle\hspace{50mm}+\frac{\alpha^3}{60\pi}\left[1-\textrm{erf}\left(\frac{\alpha\varepsilon}{2}\right)\right],\hfill r=0 \end{array}\right.\\\nonumber\\
   Q_1^{\varepsilon}(r)&=&
   \left\{    \begin{array}{ll}
\displaystyle\frac{2e^{-r^2/\varepsilon^2}}{\pi^{3/2}\alpha^2\varepsilon^5}\left(1-e^{-\alpha^2\varepsilon^2/4}\right)-\frac{1}{2\pi^{3/2}\varepsilon r^2}e^{-r^2/\varepsilon^2}+\frac{1}{4\pi r^3}\textrm{erf}\left(\frac{r}{\varepsilon}\right), & r> 0 \\\\
\displaystyle\frac{1}{3\pi^{3/2}\alpha^2\varepsilon^5}\left(6+\alpha^2\varepsilon^2-6e^{-\alpha^2\varepsilon^2/4}\right), & r=0
\end{array}      \right.\\\nonumber\\
   Q_2^{\varepsilon}(r)&=&
      \left\{    \begin{array}{ll}
\displaystyle-\frac{1}{2}H_1^{\varepsilon}(r)-\frac{r^2}{2}H_2^{\varepsilon}(r), & r> 0 \\
\displaystyle-\frac{e^{-\alpha^2\varepsilon^2/4}}{6\pi^{3/2}\varepsilon}+\frac{1}{3\pi^{3/2}\alpha^2\varepsilon^3}\left(-1+e^{-\alpha^2\varepsilon^2/4}\right)+\frac{\alpha}{12\pi}\left[1-\textrm{erf}\left(\frac{\alpha\varepsilon}{2}\right)\right], & r=0 
\end{array}      \right.\\\nonumber\\
   D_1^{\varepsilon}(r)&=&
         \left\{    \begin{array}{ll}
\displaystyle\phi_{\varepsilon}(r)-Q_1^{\varepsilon}(r), & r> 0 \\
\displaystyle-\frac{4}{\pi^{3/2}\alpha^2\varepsilon^5}e^{-\alpha^2\varepsilon^2/4}+\frac{2(6+\alpha^2\varepsilon^2)}{3\pi^{3/2}\alpha^2\varepsilon^5}, & r=0 
\end{array}      \right.\\
   D_2^{\varepsilon}(r)&=& 
            \left\{    \begin{array}{ll}
\displaystyle\frac{4}{\pi^{3/2}\alpha^2\varepsilon^7}e^{-r^2/\varepsilon^2}\left(1-e^{-\alpha^2\varepsilon^2/4}\right)-\frac{3\varepsilon^2+2r^2}{2\pi^{3/2}\varepsilon^3r^4}e^{-r^2/\varepsilon^2}+\frac{3}{4\pi r^5}\textrm{erf}\left(\frac{r}{\varepsilon}\right), & r> 0 \\\\
\displaystyle\frac{2}{5\pi^{3/2}\alpha^2\varepsilon^7}\left(10+\alpha^2\varepsilon^2-10e^{-\alpha^2\varepsilon^2/4}\right), & r=0. \label{D2blob}
\end{array}      \right.
  \end{eqnarray} 

     \subsection{Comparing Blob Functions \& Methods}\label{appcompare}
\begin{figure}
 \hspace{1.75cm}{\bf(a)}\hspace{5.5cm}{\bf(b)}\\ \vspace{-.3cm}
 \centering
\includegraphics[width=.41\linewidth]{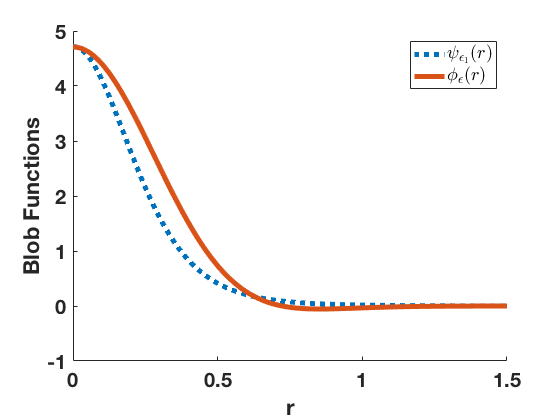}
\includegraphics[width=.41\linewidth]{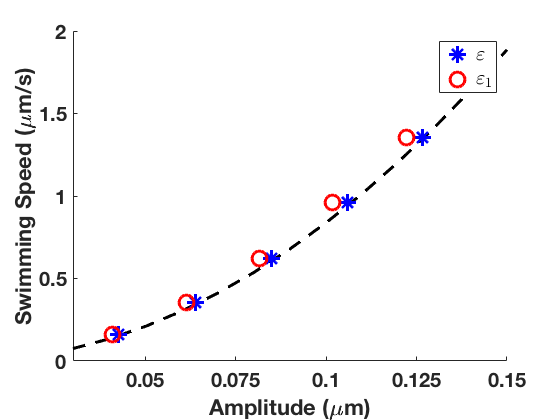}
\caption{{(a) Blob function $\psi_{\varepsilon_1}$ in Eq.~(\ref{BloBFunction1}) and $\phi_{\varepsilon}$ in Eq.~(\ref{NganBlobFunction}) are plotted for the case of $\alpha\to0$. The regularization parameter $\varepsilon_1=1$ and $\varepsilon=0.909$}. (b) Comparing solutions found by choosing a blob function (using $\varepsilon=6.363\triangle s$) and by regularizing the fundamental solution (Section \ref{KRRegularizedApproach}, with  $\varepsilon_1=7\triangle s$) for the case of a planar bending swimmer with $\sigma=350$.}\label{LimitBloBs}
\end{figure}

The solution to the fluid flow does depend on the particular choice of blob function. The blob function $\psi_{\varepsilon_1}$ in Eq.~(\ref{BloBFunction1}) has algebraic decay in the range between $r^{-7}$ and $r^{-5}$ as $r\to\infty$ whereas $\phi_{\varepsilon}$ in Eq.~(\ref{NganBlobFunction}) decays exponentially, independent of the choice of $\alpha$. To compare these two blob functions fairly, we match their limits and obtain the relation $\varepsilon=0.909\varepsilon_1$ so that they agree at $r=0$. We plot the functions in Eq.~(\ref{BloBFunction1}) and Eq.~(\ref{NganBlobFunction}) for the case when $\alpha\to0$ with $\varepsilon_1=1$ and $\varepsilon=0.909$. Figure \ref{LimitBloBs} shows the differences for $0<r<1$ and the agreement in the far-field as the solutions decay to zero. Here, the evaluation at a point force at $r=0$ will give the same value for the flow, but will given slightly different results off of the structure.
 
 The computational swimming speeds for planar bending (using the setup in Section \ref{planarsection}) is explored using the two different approaches presented in Section \ref{TwoApproaches} and detailed in Appendix \ref{option1app}--\ref{option2app}. The numerical results are obtained from choosing the blob function (Section \ref{KRBlobApproach}) with regularization parameter $\varepsilon=6.363\triangle s$ and regularizing the fundamental solutions (Section \ref{KRRegularizedApproach}) with $\varepsilon_1=7\triangle s$. In figure \ref{LimitBloBs}(b), for the case of $L=40$, numerical results from both approaches match with the asymptotic swimming speeds. The figure also shows that for this particular case of planar bending, the solutions obtained from the method where a blob function is chosen first yields slightly better agreement in terms of the swimming speeds. In addition, for this method, the achieved amplitude $b$ is closer to the preferred amplitude. Thus, computational results that are presented in the results section will be for the case of choosing a particular blob function first (Section \ref{KRBlobApproach}). 

   \section{Linear and Angular Velocity as $\alpha\to0$}\label{limits}
 The equations of the linear and angular velocity of a KR in a Brinkman fluid are given in (\ref{uvel_B}) and (\ref{wvel_B}), respectively. 
When $\alpha\to0$, the linear and angular velocity become
           \begin{eqnarray*}
 \mu{\bf u}({\bf x})&=&{\bf f}_cH_1^{\varepsilon}(r)+({\bf f}_c\cdot{\bf \hat{x}}){\bf \hat{x}}H_2^{\varepsilon}(r)+\frac{1}{2}({\bf m}_c\times{\bf \hat{x}})Q_1^{\varepsilon}(r),\\
 \mu{\boldsymbol \omega}({\bf x})&=&\frac{1}{2}({\bf f}_c\times{\bf \hat{x}})Q_1^{\varepsilon}(r)+\frac{1}{4}\left[{\bf m}_cD_1^{\varepsilon}(r)+({\bf m}_c\cdot{\bf \hat{x}}){\bf \hat{x}}D_2^{\varepsilon}(r)\right].
   \end{eqnarray*}  
In order to show that these equations exactly approach the regularized Stokes flow as $\alpha\to0$, we next need to show that the Brinkman regularized coefficients $H_i^{\varepsilon}$, $Q_i^{\varepsilon}$, and $D_i^{\varepsilon}$ for $i=1,2$ approach the Stokes regularized coefficients for a specified blob function. As mentioned in Section \ref{TwoApproaches}, the coefficients can be calculated in two different ways, which are detailed below.
      
      \subsection{Regularizing the Fundamental Solutions}\label{MRBrinman1}%
 When $\alpha\to0$, the coefficients $H_1^{\varepsilon}(r), H_2^{\varepsilon}(r), Q_1^{\varepsilon}(r), D_1^{\varepsilon}(r)$, and $D_2^{\varepsilon}(r)$ given in (\ref{H1singular})--(\ref{D2reg}) match exactly with those derived in \cite{Olson13}. That is, when $\alpha\to0$, the solution corresponds to the linear and angular velocity of a KR in Stokes flow when using the blob function $\psi_{\varepsilon}(r)=\frac{15\varepsilon^4}{8\pi(r^2+\varepsilon^2)^{7/2}}$.
   
   \subsection{Selecting a Blob Function}\label{MRBrinman2}
When $\alpha\to0$, the coefficients $H_1^{\varepsilon}(r), H_2^{\varepsilon}(r), Q_1^{\varepsilon}(r), D_1^{\varepsilon}(r)$, and $D_2^{\varepsilon}(r)$ give in (\ref{H1blob})--(\ref{D2blob}) are
   \begin{small}
  \begin{eqnarray*}
 H_1^{\varepsilon}(r)&=& \left\{
    \begin{array}{ll}
\displaystyle\frac{1}{4\pi^{3/2}\varepsilon}e^{-r^2/\varepsilon^2}+\frac{1}{8\pi}\textrm{erf}\left(\frac{r}{\varepsilon}\right), & r> 0 \\
\displaystyle\frac{1}{2\pi^{3/2}\varepsilon},& r=0
    \end{array} \right.\\\nonumber\\
H_2^{\varepsilon}(r)&=&\left\{
    \begin{array}{ll}
\displaystyle\frac{1}{8\pi^{3/2}r^3}\left[-\frac{2r}{\varepsilon}e^{-r^2/\varepsilon^2}+\sqrt{\pi}\textrm{erf}\left(\frac{r}{\varepsilon}\right)\right], & r> 0 \\
\displaystyle\frac{1}{6\pi^{3/2}\varepsilon^3},& r=0
    \end{array} \right.\\\nonumber\\
   Q_1^{\varepsilon}(r)&=&   \left\{
    \begin{array}{ll}
\displaystyle\frac{e^{-r^2/\varepsilon^2}}{4\pi^{3/2}r^3\varepsilon^3}\left(-2r\varepsilon^2+2r^3\right)+\frac{1}{4\pi r^3}\textrm{erf}\left(\frac{r}{\varepsilon}\right), & r> 0 \\
\displaystyle\frac{5}{6\pi^{3/2}\varepsilon^3}, & r=0
    \end{array} \right.\\\nonumber\\
   D_1^{\varepsilon}(r)&=&   \left\{
    \begin{array}{ll}
\displaystyle\frac{-e^{-r^2/\varepsilon^2}}{4\pi^{3/2}r^5\varepsilon^3}\left(-2r\varepsilon^4-8r^3\varepsilon^2+4r^5\right)-\frac{1}{4\pi r^3}\textrm{erf}\left(\frac{r}{\varepsilon}\right), & r> 0 \\
\displaystyle\frac{5}{3\pi^{3/2}\varepsilon^3}, & r=0 \\
    \end{array} \right.\\\nonumber\\
   D_2^{\varepsilon}(r)&=&     \left\{
    \begin{array}{ll}
\displaystyle\frac{e^{-r^2/\varepsilon^2}}{4\pi^{3/2}r^5\varepsilon^3}\left(-6r\varepsilon^4-4r^3\varepsilon^2+4r^5\right)+\frac{3}{4\pi r^3}\textrm{erf}\left(\frac{r}{\varepsilon}\right), & r> 0 \\
\displaystyle\frac{7}{5\pi^{3/2}\varepsilon^5}, & r=0. 
    \end{array} \right.\\\nonumber
  \end{eqnarray*} 
   \end{small}
This means that as $\alpha\to0$, the solution for the linear and angular velocity of a KR corresponds to regularized Stokes flow when using the blob function $\phi_{\varepsilon}(r)=\frac{1}{\pi^{3/2}\varepsilon^3}\left(\frac{5}{2}-\frac{r^2}{\varepsilon^2}\right)e^{-r^2/\varepsilon^2}$.

\section{Right-Handed Helix Coefficients}\label{HelixDerived}
\noindent We consider the right-handed helix parameterized as
\begin{eqnarray}
{\bf r}_h(t)=\{r_h\cos t, r_h\sin t, pt\},\label{Eq1.1}
\end{eqnarray}
where $r_h$ is the radius of the helix and $p$ is the reduced pitch of the helix (the actual pitch is $2\pi p$ \citep{goriely1997nonlinear}). The task is to write the helix in terms of the arc length $s$ and determine the relations among the radius $r_h$, the pitch $p$, the curvature $\kappa$, and the torsion $\tau$. The arc length from $0$ to $t$ is
  \begin{eqnarray}
s(t)=\int_0^{t}\left\|{\bf r}'_h(t)\right\|dt=\int_0^t\sqrt{r_h^2+p^2}dt=t\sqrt{r_h^2+p^2}.
\end{eqnarray}
Then, 
  \begin{eqnarray}
t=\frac{s}{\sqrt{r_h^2+p^2}}=\xi s,
\end{eqnarray}
where $\xi=1/\sqrt{r_h^2+p^2}$. The helix can be rewritten in terms of arc length as 
\begin{eqnarray}
{\bf r}_h(s)=\{r_h\cos \xi s, r_h\sin \xi s, p\xi s\},
\end{eqnarray}
where the curvature $\kappa$ of the helix is determined as 
  \begin{eqnarray}
\kappa=\left\|{\bf r}_h''(s)\right\|=\frac{r_h}{r_h^2+p^2}.
\end{eqnarray}
To determine the torsion $\tau$, we consider the Frenet frame with tangent vector ${\bf T}(s)$, normal vector ${\bf N}(s)$, and binormal vector ${\bf B}(s)$, which satisfy the following \citep{chouaieb2004kirchhoff}:
  \begin{eqnarray}
{\bf T}'(s)&=&\kappa{\bf N}(s)\label{Frenet1.1},\\
{\bf N}'(s)&=&\tau {\bf B}(s)-\kappa{\bf T}(s)\label{Frenet1.2},\\
{\bf B}'(s)&=&-\tau{\bf N}(s).\label{Frenet1.3}
\end{eqnarray}
We can also write ${\bf T}(s)$, ${\bf N}(s)$, and ${\bf B}(s)$ as
  \begin{eqnarray}
{\bf T}(s)&=&{\bf r}_h'(s)=\{-r_h\xi\sin\xi s, r_h\xi\cos\xi s, p\xi\},\label{tangentvec}\\
{\bf N}(s)&=&\frac{{\bf r}_h''(s)}{\|{\bf r}_h''(s)\|}=\{-\cos\xi s, \sin\xi s, 0\},\label{normalvec}\\
{\bf B}(s)&=&{\bf T}(s)\times{\bf N}(s)=\{p\xi\sin\xi s, -p\xi\cos\xi s, r_h\xi\}\label{binormalvec}.
\end{eqnarray}
To find the expression for $\tau$, we need to calculate ${\bf B}'(s)$. From (\ref{binormalvec}) and (\ref{Frenet1.1}), we have 
  \begin{eqnarray}
{\bf B}'(s)&=&\left({\bf T}(s)\times{\bf N}(s)\right)'={\bf T}'(s)\times{\bf N}(s)+{\bf T}(s)\times{\bf N}'(s),\nonumber\\
&=&\kappa{\bf N}(s)\times {\bf N}(s)+{\bf T}(s)\times{\bf N}'(s),\nonumber\\
&=&\bf{T}(s)\times{\bf N}'(s),\label{bionormalPrime}
\end{eqnarray}
where ${\bf N}'(s)=\xi\{\sin\xi s, -\cos\xi s, 0\}$. Substituting (\ref{tangentvec}) into (\ref{bionormalPrime}), we arrive at
  \begin{eqnarray}
{\bf B}'(s)&=&\frac{1}{r_h^2+p^2}\{p\cos\xi s, p\sin\xi s, 0\},\nonumber\\
&=&-\frac{p}{r_h^2+p^2}\{-\cos\xi s, -\sin\xi s, 0\},\nonumber\\
{\bf B}'(s)&=&-\frac{p}{r_h^2+p^2}{\bf N}(s).\label{binormalfinal}
\end{eqnarray}
Comparing (\ref{binormalfinal}) and (\ref{Frenet1.3}), we have that $\tau=\frac{p}{r_h^2+p^2}$. The radius and the pitch in terms of the curvature and the torsion are 
  \begin{eqnarray}
r_h=\frac{\kappa}{\kappa^2+\tau^2}, \hspace{5mm}p=\frac{\tau}{\kappa^2+\tau^2}.
\end{eqnarray}
The right-handed helix can now be written in terms of the curvature and torsion as 
 \begin{eqnarray*}
{\bf r}_h(s)=\left\{\frac{\kappa}{\kappa^2+\tau^2}\cos\left(\sqrt{\kappa^2+\tau^2}s\right), \frac{\kappa}{\kappa^2+\tau^2}\sin\left(\sqrt{\kappa^2+\tau^2}s\right), \frac{\tau}{\sqrt{\kappa^2+\tau^2}}s\right\}.
\end{eqnarray*}
Furthermore, if we let $\theta$ be the helix angle between the tangent vector and the $z$-axis, then $\theta$ relates to $r_h$ and $p$ as follows,
  \begin{eqnarray}
\cos\theta=\frac{{\bf T}(s)\cdot{\bf e}_z}{\|{\bf T}(s)\|\|{\bf e}_z\|}={\bf T}(s)\cdot{\bf e}_z=p\xi \hspace{5mm} \textrm{ and}  \hspace{5mm} \sin\theta=r_h\xi,
\end{eqnarray}
where ${\bf e}_z$ is the unit vector in the $z$-direction.

\subsection{Material Frame vs. Frenet Frame}
\noindent Consider an inextensible, unshearable, and uniform rod ${\bf r}_h(s)$ discretized at the centerline, where $s$ is arc length such that $s\in[0,L]$ for rod length $L$. The corresponding orthonormal triads of the rod $\{{\bf D}^1(s), {\bf D}^2(s), {\bf D}^3(s)\}$ are defined such that ${\bf D}^1(s)$ is perpendicular to ${\bf D}^3(s)$ and ${\bf D}^2(s)={\bf D}^3(s)\times{\bf D}^1(s)$. This director basis relates to the Frenet basis consisting of normal ${\bf N}(s)$, binormal ${\bf B}(s)$, and tangent vector ${\bf T}(s)={\bf D}^3(s)$ as \citep{djurivckovic2013twist}
  \begin{eqnarray}
{\bf D}^1(s)&=&\cos\phi{\bf N}(s)+\sin\phi{\bf B}(s),\nonumber\\
{\bf D}^2(s)&=&-\sin\phi{\bf N}(s)+\cos\phi{\bf B}(s),\label{FrenetMaterial}\\
{\bf D}^3(s)&=&{\bf T}(s),\nonumber
\end{eqnarray}
where $\phi=\phi(s)$ is the rotation angle depending on the arc length $s$. There exists a rotational vector (a strain-twist vector) ${\bf K}=\Omega_1{\bf D}^1(s)+\Omega_2{\bf D}^2(s)+\Omega_3{\bf D}^3(s)$ such that
  \begin{eqnarray}
({\bf D}^1(s))'={\bf K}\times{\bf D}^1(s), \hspace{5mm}({\bf D}^2(s))'={\bf K}\times{\bf D}^2(s),\hspace{5mm}({\bf D}^3(s))'={\bf K}\times{\bf D}^3(s),
\end{eqnarray}
for 
  \begin{eqnarray}
\Omega^2=\Omega_1^2+\Omega_2^2,\label{Eq11}
\end{eqnarray}
where $\Omega$ is the intrinsic curvature, $\Omega_1$ is the geodesic curvature, $\Omega_2$ is the normal curvature, and $\Omega_3$ is the twist \citep{dineen1998multivariate}. Then,
  \begin{eqnarray}
({\bf D}^1(s))'&=&\Omega_3{\bf D}^2(s)-\Omega_2{\bf D}^3(s),\nonumber\\
({\bf D}^2(s))'&=&-\Omega_3{\bf D}^1(s)+\Omega_1{\bf D}^3(s), \label{Eq11}\\
({\bf D}^3(s))'&=&\Omega_2{\bf D}^1(s)-\Omega_1{\bf D}^2(s).\nonumber
\end{eqnarray}
Rewriting the system (\ref{FrenetMaterial}) using  Eqs.~(\ref{tangentvec})--(\ref{binormalvec}), ${\bf N}$ and ${\bf B}$  are in terms of ${\bf D}^1(s)$ and ${\bf D}^2(s)$ as follows
  \begin{eqnarray}
{\bf N}(s)=\cos\phi{\bf D}^1(s)-\sin\phi{\bf D}^2(s), \hspace{5mm}{\bf B}(s)=\sin\phi{\bf D}^1(s)+\cos\phi{\bf D}^2(s).\label{Eq13}
\end{eqnarray}
Taking the first derivative of ${\bf B}$ in (\ref{Eq13}) and using (\ref{Eq11})$-$(\ref{Eq13}), we have
  \begin{eqnarray*}
{\bf B}'(s)&=&\phi'\left[\cos\phi{\bf D}^1(s)-\sin\phi{\bf D}^2(s)\right]+\left[\sin\phi({\bf D}^1)'(s)+\cos\phi({\bf D}^2)'(s)\right],\\
&=&\phi'{\bf N}(s)+\Omega_3\left(\sin\phi{\bf D}^2(s)-\cos\phi{\bf D}^1(s)\right)+(-\Omega_2\sin\phi+\Omega_1\cos\phi){\bf T}(s),\\
&=&(\phi'-\Omega_3){\bf N}(s)+(-\Omega_2\sin\phi+\Omega_1\cos\phi){\bf T}(s).
\end{eqnarray*}
Comparing with ${\bf B}'(s)=-\tau{\bf N}$(s) in (\ref{Frenet1.3}), 
  $\Omega_3=\phi'+\tau$
and
  $-\Omega_2\sin\phi+\Omega_1\cos\phi=0$  or  $\Omega_1\cos\phi=\Omega_2\sin\phi$. 
Then, with (\ref{Eq11}) we have
  \begin{eqnarray*}
\tan^2\phi&=&\frac{\Omega_1^2}{\Omega_2^2}=\frac{\Omega^2-\Omega_2^2}{\Omega_2^2},\\
\Omega_2^2&=&\frac{\Omega^2}{1+\tan^2\phi}=\Omega^2\cos^2\phi,
\end{eqnarray*}
or,
$$\Omega_2=\Omega\cos\phi, \hspace{5mm}\textrm{and}\hspace{5mm}\Omega_1=\Omega\sin\phi.$$
Therefore, the strain-twist vector (or the rotating vector) becomes
  \begin{eqnarray}
{\bf K}=\{\Omega\sin\phi, \Omega\cos\phi, \phi'+\tau\}.\label{Eq17}
\end{eqnarray}
If the preferred curvature of the helical bending wave has no twisting component, then 
  \begin{eqnarray}
\Omega_3=\phi'+\tau=0.
\end{eqnarray}
Thus, the rotating angle relates to the torsion $\tau$ as 
  \begin{eqnarray}
\phi=-\int_0^s\tau ds.\label{phirotint}
\end{eqnarray}
\subsection{Helical Bending Wave}
\noindent If the torsion $\tau$ is assumed to be constant, then the rotating angle (\ref{phirotint}) is $\phi=-\tau s$. Thus, 
  \begin{eqnarray}
{\bf K}=\{-\Omega\sin(\tau s), \Omega\cos(\tau s), 0\},
\end{eqnarray}
and
  \begin{eqnarray}
\left[
\begin{array}{ccccc}
{\bf D}^1(s)\\
{\bf D}^2(s)\\
{\bf D}^3(s)\\
\end{array}  \right]=
 \left[
\begin{array}{ccccc}
 \cos(\tau s) & -\sin(\tau s) & 0 \\
\sin(\tau s) & \cos(\tau s) & 0 \\
0& 0 &1  \\
\end{array}  \right]
 \left[\begin{array}{ccccc}
{\bf N}\\
{\bf B}\\
{\bf T}\\
\end{array}  \right].
\end{eqnarray}


\end{document}